\documentclass[journal,draftcls,onecolumn]{IEEEtran}
\usepackage{cite}

\usepackage{amssymb}
\ifCLASSINFOpdf
   \usepackage[pdftex]{graphicx}
\else
   \usepackage[dvips]{graphicx}
\fi

\usepackage[cmex10]{amsmath}
\usepackage{mdwmath}
\usepackage[lined,boxed,commentsnumbered, ruled]{algorithm2e}

\usepackage{array}

\usepackage{multirow}  % multiple row in table
\usepackage{array} % fixed talbe
\newcommand{\PreserveBackslash}[1]{\let\temp=\\#1\let\\=\temp}
\newcolumntype{C}[1]{>{\PreserveBackslash\centering}p{#1}}
\newcolumntype{R}[1]{>{\PreserveBackslash\raggedleft}p{#1}}
\newcolumntype{L}[1]{>{\PreserveBackslash\raggedright}p{#1}}
\usepackage{chemarrow}

\ifCLASSOPTIONcompsoc 
\usepackage[caption=false,font=normalsize,labelfont=sf,textfont=sf]{subfig}
\else
\usepackage[caption=false,font=footnotesize]{subfig}
\fi

\usepackage{stfloats}
\usepackage{extarrows}
\allowdisplaybreaks

\begin{document}
\title{Time-Domain {\it N}-continuous GFDM}
%nonbreaking spaces ( ~ ) LaTeX will not break

\author{Peng~Wei}%, Yue~Xiao, Shaoqian~Li % <-this % stops a space
\maketitle

\begin{abstract}
\boldmath
Generalized frequency division multiplexing (GFDM) has been a candidate multicarrier scheme in the 5th generation cellular networks for its flexibility of transmitter filter in time and frequency. However, for the circularly shaped transmitter filter, GFDM provides limited performance gain of sidelobe suppression. In this paper, we propose a scheme, called time-domain {\it N}-continuous GFDM (TD-NC-GFDM), to reduce the discontinuities caused by the GFDM transmitter filter and achieve promising sidelobe suppression gain. Based on time-domain {\it N}-continuous orthogonal frequency devision multiplexing (TD-NC-OFDM), TD-NC-GFDM signal can be obtained by superposing a smooth signal in the time domain. The smooth signal is linearly combined by basis signals in a new basis set related to GFDM transmitter waveform. To eliminate the interference caused by the smooth signal, two solutions are proposed. Firstly, a signal recovery algorithm for reception is adopted at the cost of high complexity.  Thus, secondly, to simplify the TD-NC-GFDM receiver, a low-interference TD-NC-GFDM is proposed by redesigning the basis signals. A soft truncation of the basis signals in TD-NC-GFDM is given to design the basis signals in the low-interference TD-NC-GFDM. Then, the smooth signal is aligned with the beginning of the GFDM symbol and is added in the front part of the GFDM symbol. Moreover, for a big number of GFDM subsymbols, theoretical analysis proves that the signal-to-interference ratio (SIR) in TD-NC-GFDM is much higher than that in TD-NC-OFDM. Simulation results shows that TD-NC-GFDM can obtain significant sidelobe suppression performance as well as the low-interference TD-NC-GFDM, which can achieve the same BER performance as the original GFDM.
\end{abstract}


\IEEEpeerreviewmaketitle

%%%%%%%%%%%%%%%%%%%%%%%%%%%%%%%%%%%%%%%%%%%%%%%%%%%%%%%%%%%%%%%%%%%%%%%%%%%%%%%%%%%%%%%%%%%%%%%%%%%%%%%%%%%%%%%%
\section{Introduction}

Recently, generalized frequency division multiplexing (GFDM) has been paid more attention in the 5th generation cellular networks for its flexible pulse shaping in time and frequency \cite{Ref1,Ref2,Ref3,Ref4,Ref5,Ref6,Ref7,Ref8}. Since the GFDM transmitter filter has small sidelobes, GFDM can obtain lower spectral leakage than orthogonal frequency devision multiplexing (OFDM) \cite{Ref1,Ref6}. Unfortunately, the circularly shaped waveforms of the transmitter filter \cite{Ref1, Ref8} make the GFDM signal discontinuous, which causes the spectral leakage and reduces the spectral efficiency in GFDM systems. Thus, to achieve the requirement of high spectral efficiency in the 5th generation communications \cite{Ref3}, a significant sidelobe suppression performance is required.

To suppress the spectral leakage, the windowing technique has been adopted to GFDM signal in \cite{Ref1}. However, the extended guard interval used to avoid signal distortion results in a reduction in spectral efficiency. Meanwhile, as analyzed in \cite{Ref8}, a ramp-up at the beginning of each GFDM symbol and a ramp-down at the end of each GFDM symbol will be caused. There has been many sidelobe suppression techniques for OFDM signal \cite{Ref9,Ref10,Ref11,Ref12,Ref13,Ref14,Ref15,Ref16,Ref17,Ref18,Ref19,Ref20,Ref21,Ref22} which can be applied to GFDM signal. Cancellation carriers in \cite{Ref9,Ref10} consume extra power and cause performance loss in terms of signal-to-noise ratio (SNR) and data rate. Interference cancellation methods \cite{Ref11,Ref12} lead to severe inter symbol interference (ISI) among multiple symbols while requiring large-scale matrix operations for a wide optimization frequency band. The precoding methods \cite{Ref13,Ref14,Ref15} have a promising sidelobe suppression performance at the cost of high complexity at the transmitter \cite{Ref13} and receiver \cite{Ref13,Ref14,Ref15}. {\it N}-continuous OFDM (NC-OFDM) techniques \cite{Ref16,Ref17,Ref18,Ref19,Ref20,Ref21,Ref22} make the OFDM signal and its first {\it N} derivatives continuous, termed as {\it N}-continuous, to achieve a considerable sidelobe suppression performance. Conventional NC-OFDM \cite{Ref16} utilizes a precoder to obtain {\it N}-continuous signal at the price of increased complexity and high interference. To optimize the precoder in NC-OFDM \cite{Ref16}, the technique in \cite{Ref17} nulls the spectrum at several chosen frequencies at the expense of the bit-to-error ratio (BER) performance degradation. The precoder in \cite{Ref18} improves the BER performance compared to that in \cite{Ref17}, but results in data rate loss. Aiming at the low-complexity signal recovery for NC-OFDM signal, several techniques have been investigated in \cite{Ref19,Ref20}. However, the technique in \cite{Ref19} causes a significant increase in transmitter complexity when searching the optimal data sequence with minimum interference, and the method in \cite{Ref20} also increases transmitter complexity in the calculation of cancellation tones. In \cite{Ref21,Ref22}, time-domain {\it N}-continuous OFDM (TD-NC-OFDM) is proposed to reduce the precoder complexity of the conventional NC-OFDM while maintaining as good sidelobe suppression as the conventional NC-OFDM, but still causes high interference.

In this paper, in order to achieve significant sidelobe suppression, based on \cite{Ref21,Ref22}, a time-domain  {\it N}-continuous GFDM (TD-NC-GFDM) is proposed by adding a smooth signal into the GFDM symbol. The smooth signal is a linear combination of the basis signals in a new basis set, which is designed by the GFDM transmitter filter and its high-order derivatives. To reduce the interference caused by the smooth signal, we adopt two solutions. The first one is an iterative signal recovery algorithm for the received TD-NC-GFDM signal by a calculated decoding matrix, which increases the receiver complexity. To avoid the complexity of the signal recovery algorithm, in the second solution, a low-interference TD-NC-GFDM is presented by redesigning the basis signals. In this paper, a redesign of the basis signals is to truncate the aforementioned basis signals in TD-NC-GFDM by a smooth window function. After the linear combination of the redesigned basis signals, the resulted smooth signal is aligned with the beginning of the GFDM symbol and added in the front part of the GFDM symbol. Furthermore, for the truncation design of the basis signals, the low-interference scheme has lower implementation complexity than TD-NC-GFDM. Additionally, it is proved that when the number of GFDM subsymbols in each GFDM symbol block is bigger than 1,  the signal-to-interference ratio (SIR) in TD-NC-GFDM is much higher than that in TD-NC-OFDM. Simulation results shows that TD-NC-GFDM and its low-interference scheme can obtain much more rapid sidelobe decaying than TD-NC-OFDM, and the low-interference TD-NC-GFDM can obtain the same BER performance as the original GFDM, which has the close BER performance to OFDM for a small roll-off factor of the transmitter filter. 

The remainder of the paper is organized as follows. In Section II, conventional GFDM transmitter and receiver are introduced. Section III presents the TD-NC-GFDM at the transmitter followed by the signal recovery algorithm at the receiver, gives the low-interference TD-NC-GFDM, and analyzes the SIR of the TD-NC-GFDM signal. Simulation results follows in Section IV. Finally, in Section V, this paper is concluded.

%%%%%%%%%%%%%%%%%%%%%%%%%%%%%%%%%%%%%%%%%%%%%%%%%%%%%%%%%%%%%%%%%%%%%%%%%%%%%%%%%%%%%%%%%%%%%%%%%%%%%%%%%%%%%%%%
\section{GFDM system}

%=========================================================================================================================================================================
\subsection{GFDM Transmitter}

\begin{figure}[ht]
\centering
\includegraphics[width=6in]{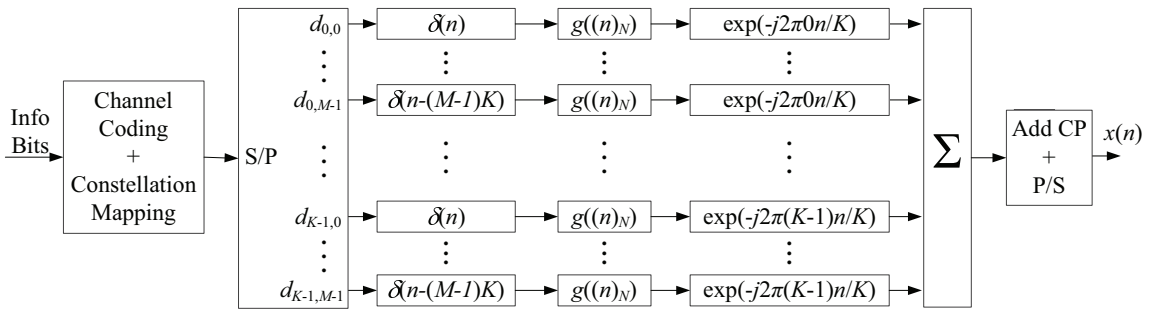}
\caption{Block diagram of GFDM transmitter.}
\label{fig:fig1}
\end{figure}

In GFDM transmitter, bit streams are first modulated to complex symbols $d_{i, k, m}$ that are divided into sequences of \emph{KM} symbols long in the {\it i}th GFDM symbol. Each sequence (as a vector) $\mathbf{d}_i=[\mathbf{d}^{\rm T}_{i,0}, \mathbf{d}^{\rm T}_{i,1}, \ldots, \mathbf{d}^{\rm T}_{i,M-1}]^{\rm T}$ with $\mathbf{d}_{i, m}=[d_{i, 0,m}, d_{i,1,m}, \ldots, d_{i,K-1,m}]^{\rm T}$, $m=0, 1, \ldots, M-1$, is spread on \emph{K} subcarriers in \emph{M} time slots. Therein, $d_{i,k,m}$ is the transmitted data on the \emph{k}th subcarrier in the \emph{m}th subsymbol of the {\it i}th GFDM block. The data symbols are taken from an independent and identically distributed (i.i.d.) process with the zero mean and the unit variance. Each $d_{i,k,m}$ is transmitted with a pulse shaping filter \cite{Ref1}
\begin{equation}
\label{Eqn1}
g_{k,m}(n)=g\left((n-mK)_N\right)e^{-j2\pi\frac{k}{K}n},
\end{equation}
where the signal sample index is $n=0,1,\ldots, N-1$ with $N=KM$, $(\cdot)_N$ denotes the modulo of \emph{N}, and $g(n)$ is a prototype filter whose time and frequency shifts by $k$ and $m$ are $g_{k,m}(n)$. By the superposition of all the filtered $d_{i,k,m}$, the GFDM transmit signal is
\begin{equation}
\label{Eqn2}
x_i(n)=\sum\limits^{K-1}_{k=0}{\sum\limits^{M-1}_{m=0}{d_{i,k,m}g_{k,m}(n)}},
\end{equation}
and its implementation is shown in Fig. \ref{fig:fig1}  that is the same as the transmitter shown in \cite{Ref1}. Eq. \eqref{Eqn2} can be rewritten in the matrix form as
\begin{equation}
\label{Eqn3}
\mathbf{x}_i=\mathbf{A}\mathbf{d}_i,
\end{equation}
where $\mathbf{A}$ is a $N\times N$ transmitter matrix \cite{Ref1}, which is given by
\begin{equation}
\label{Eqn4}
\mathbf{A}=\left[\mathbf{g}_{0,0}  \;  \cdots  \;  \mathbf{g}_{K-1,0} \;  \; \mathbf{g}_{0,1}  \; \cdots  \; \mathbf{g}_{K-1,M-1}  \right]
\end{equation}
with $\mathbf{g}_{k,m}=[g_{k,m}(0), g_{k,m}(1), \ldots, g_{k,m}(N-1)]^{\rm T}$.

Lastly, on the transmitter side, a cyclic prefix (CP) of $N_{\rm cp}$ samples is appended to produce $\mathbf{x}_i$. In this case, the transmitted GFDM signal in the time range $(-\infty, +\infty)$ is expressed as
\begin{equation}
\label{Eqn5}
x(n)=\sum\limits^{+\infty}_{i=-\infty}{\sum\limits^{K-1}_{k=0}{\sum\limits^{M-1}_{m=0}{d_{i,k,m}g_{k,m}(n-i(N+N_{\rm cp}))}}}.
\end{equation}

%=========================================================================================================================================================================
\subsection{GFDM Receiver}

At the receiver, the {\it i}th received GFDM signal is
\begin{equation}
\label{Eqn6}
y_i(n)=h_i(n)* x_i(n)+n_i(n),
\end{equation}
where $*$ denotes the linear convolution operation, $h_i(n)$ is the channel response in the time domain, and $n_i(n)$ is the AWGN noise with zero mean and variance $\sigma^2$. 

Assuming that perfect synchronization and long enough CP against the maximum channel delay are implemented, after the CP is removed, Eq. \eqref{Eqn6} can be expressed as
\begin{equation}
\label{Eqn7}
\mathbf{y}_i=\mathbf{H}_i \mathbf{x}_i+\mathbf{n}_i,
\end{equation}
where $\mathbf{y}_i=[y_i(0), y_i(1), \ldots, y_i(N-1)]^{\rm T}$, $\mathbf{n}_i=[n_i(0), n_i(1), \ldots, n_i(N-1)]^{\rm T}$, and $\mathbf{H}_i$ denots the $N\times N$ channel matrix which is a circular convolution matrix.

According to \cite{Ref1}, three linear GFDM demodulators have been considered, which are matched filter (MF), zero-forcing (ZF), and minimum mean square error (MMSE) receivers. Suppose that zero-forcing channel equalization is adopted.  The estimated data $\hat{\mathbf{d}}_i$ by the three receivers can be, respectively, expressed by
\begin{subequations}
\label{Eqn8}
\begin{align}
&\hat{\mathbf{d}}_{i,\rm MF}=\mathbf{A}^{\rm H} \mathbf{H}^{-1}_i \mathbf{y}_i, \\
&\hat{\mathbf{d}}_{i,\rm ZF}=\mathbf{A}^{-1} \mathbf{H}^{-1}_i \mathbf{y}_i, \\
&\hat{\mathbf{d}}_{i,\rm MMSE}=\left(\mathbf{R}_{i,\rm w} +  \mathbf{A}^{\rm H} \mathbf{H}^{\rm H}_i  \mathbf{H}_i\mathbf{A}\right)^{-1} \mathbf{A}^{\rm H} \mathbf{H}^{\rm H}_i \mathbf{y}_i,
\end{align}
\end{subequations}
where $\mathbf{R}_{i,\rm w}$ is the covariance matrix of the channel noise.

Finally, the data $\hat{\mathbf{d}}_i$ are demapped to a sequence of bits by a decoder. 

For the flexibility of GFDM transmitter filter, a GFDM signal has the lower spectral leakage than the rectangularly pulsed OFDM signal \cite{Ref1}. However, for the circularly shaped transmitter filter $g_{k,m}(n)$, the spectral leakage, caused by the discontinuities between GFDM symbols, cannot be significantly suppressed. Thus, to achieve a promising suppression gain of spectral leakage, a sidelobe suppression technique, called TD-NC-GFDM, is proposed below.  

%%%%%%%%%%%%%%%%%%%%%%%%%%%%%%%%%%%%%%%%%%%%%%%%%%%%%%%%%%%%%%%%%%%%%%%%%%%%%%%%%%%%%%%%%%%%%%%%%%%%%%%%%%%%%%%%
\section{Time-Domain {\it N}-continuous GFDM}

In this section, we first formulate TD-NC-GFDM and its signal recovery algorithm. Then, a low-interference TD-NC-GFDM is proposed to reduce the receiver complexity. Lastly, the effect of the smooth signal on GFDM signal is analyzed in terms of SIR.

%===================================================================================================================================================================
\subsection{TD-NC-GFDM}
 
Based on \cite{Ref16, Ref21, Ref22}, signal's spectral leakage can be reduced by improving the signal's continuity, which is measured by the high-order derivative's continuity. According to \cite{Ref21, Ref22}, by adding a smooth signal $w_i(n)$ in each GFDM symbol, that is
\begin{equation}
\label{Eqn9}
\bar{x}_{i}(n)=x_{i}(n)+w_i(n),
\end{equation}
the transmitted GFDM signal $x(n)$ in \eqref{Eqn5} becomes {\it N}-continuous. In this case, the smooth signal $w_i(n)$ satisfies 
\begin{equation}
\label{Eqn10}
\left.w^{(v)}_i(n)\right|_{n=-N_{\rm cp}}=\left.\bar{x}^{(v)}_{i-1}(n)\right|_{n=N} - \left.x^{(v)}_{i}(n)\right|_{n=-N_{\rm cp}}
\end{equation} 
for $n\in \mathcal{N}=\{-N_{\rm cp},-N_{\rm cp}+1,\ldots,N-1\}$ in the equivalent-baseband GFDM signal with CP, where $x^{(v)}_{i}(n)$ is the {\it v}th-order derivative of $x_{i}(n)$ with $v\in\mathcal{U}_V\triangleq \{0,1,\ldots,V\}$ and the highest derivative order (HDO) $V$.

According to \cite{Ref21, Ref22}, $w_i(n)$ is generated by a linear combination, which is
 \begin{equation}
\label{Eqn11}
w_i(n)=\sum\limits^{V}_{v=0}{b_{i,v}f_v(n)},
\end{equation} 
where the calculation of the linear combination coefficients $b_{i,v}$ will be given below and the basis signal $f_v(n)$ is from a basis set $\mathcal{Q}$,
\begin{equation}
\label{Eqn12}
\mathcal{Q}=\left\{\mathbf{q}_{\tilde{v}} \left| \mathbf{q}_{\tilde{v}}=\left[ f_{\tilde{v}}(-N_{\rm cp}), f_{\tilde{v}}(-N_{\rm cp}+1), \ldots, f_{\tilde{v}}(N-1) \right]^{\rm T}, \tilde{v}\in\mathcal{U}_{2V}\right. \right\}.
\end{equation} 
Based on the prototype filter $g(n)$ in GFDM transmitter,  a design of the basis signal $f_{\tilde{v}}(n)$ is given by
\begin{equation}
\label{Eqn13}
f_{\tilde{v}}(n)=\frac{1}{N}\sum\limits^{N-1}_{l=0}{\left( j2\pi\frac{l}{N} \right)^{\tilde{v}} F_0(l) e^{j2\pi\frac{l+N_{\rm cp}}{N}n} },
\end{equation} 
where $F_0(l)$ is the $N$-point DFT of $f_0(n)$, and $f_0(n)=\sum\limits^{K-1}_{k=0}{g(n)e^{-j2\pi\frac{k}{K}n}}$. Thus, $f_{\tilde{v}}(n)$ is the $\tilde{v}$th-order derivative of $f_0(n)$ in \eqref{Eqn13}.
 
To calculate $b_{i,v}$, we first rewrite \eqref{Eqn11} in the matrix form as
 \begin{equation}
\label{Eqn14}
\mathbf{w}_i=\mathbf{Q}\mathbf{b}_{i},
\end{equation} 
where $\mathbf{Q}=[\mathbf{q}_0 \; \mathbf{q}_1 \; \cdots \; \mathbf{q}_V]$ and $\mathbf{b}_i=[b_{i,0}, b_{i,1}, \ldots, b_{i,V}]^{\rm T}$. 
Then, substituting \eqref{Eqn9} and \eqref{Eqn11} into \eqref{Eqn10}, we have
\begin{equation}
\label{Eqn15}
  \mathbf{P}_f\mathbf{b}_i=\Delta\mathbf{x}_i,
\end{equation} 
where $\mathbf{P}_f$ is a $(V+1)\times(V+1)$ symmetric matrix without correlation among its rows or columns, given as
\begin{equation}
\label{Eqn16}
\mathbf{P}_f=\! \! \begin{bmatrix}
f(-N_{\rm cp}) & f_1(-N_{\rm cp}) & \!\! \cdots \!\! & f_V(-N_{\rm cp})\\
f_1(-N_{\rm cp}) & f_2(-N_{\rm cp}) & \!\! \cdots \!\! & f_{V+1}(-N_{\rm cp})\\
\vdots & \vdots & \!\! \!\! & \vdots \\
f_V(-N_{\rm cp}) & f_{V+1}(-N_{\rm cp}) & \!\! \cdots \!\! & f_{2V}(-N_{\rm cp})
\end{bmatrix},
\end{equation}
and its inverse $\mathbf{P}^{-1}_f$ exists. Thus, the coefficients $b_{i,v}$ are solvable. The vector $$\Delta\mathbf{x}_i=\left[\bar{x}_{i-1}(N) - x_i(-N_{\rm cp}), \; \bar{x}^{(1)}_{i-1}(N)-x^{(1)}_i(-N_{\rm cp}), \; \ldots, \; \bar{x}^{(V)}_{i-1}(N)-x^{(V)}_i(-N_{\rm cp})\right]^{\rm T}$$ denotes the differences between the GFDM signals and its first \emph{V} derivatives at the adjacent point between two consecutive GFDM symbols, which can be calculated by
\begin{equation}
\label{Eqn17}
  \Delta\mathbf{x}_i= \mathbf{P}_1\bar{\mathbf{d}}_{i-1}-\mathbf{P}_2\mathbf{d}_i,
\end{equation}
where  $\bar{\mathbf{d}}_{i-1}={\mathbf{d}}_{i-1}+\mathbf{A}^{-1}\mathbf{w}_{i-1}$. Thus, Eq. \eqref{Eqn17} can be rewritten as
\begin{equation}
\label{Eqn18}
  \Delta\mathbf{x}_i= \mathbf{P}_1{\mathbf{d}}_{i-1}+\mathbf{P}_1\mathbf{A}^{-1}\mathbf{w}_{i-1}-\mathbf{P}_2\mathbf{d}_i
\end{equation}
with
\begin{equation}
\label{Eqn19}
\mathbf{P}_1=\mathbf{B}\mathbf{G},
\end{equation}
and
\begin{equation}
\label{Eqn20}
\mathbf{P}_2=\mathbf{B}\mathbf{\Phi}\mathbf{G},
\end{equation}
where
\begin{equation}
\label{Eqn21}
\mathbf{B}=\frac{1}{N} \begin{bmatrix}
1  &  1 &  \cdots   & 1 \\
j2\pi \frac{0}{N}   &    j2\pi \frac{1}{N}  &  \cdots    &  j2\pi \frac{N-1}{N} \\
\vdots &  \vdots & {} & \vdots \\
\left( j2\pi \frac{0}{N} \right)^V  & \left( j2\pi \frac{1}{N} \right)^V  &  \cdots  &  \left( j2\pi \frac{N-1}{N} \right)^V \\
\end{bmatrix},
\end{equation}
\begin{equation}
\label{Eqn22}
\mathbf{G}= \begin{bmatrix}
\mathbf{G}_{0,0}  &  \cdots &  \mathbf{G}_{K-1,0} &  \cdots  & \mathbf{G}_{0,M-1} & \cdots  & \mathbf{G}_{K-1,M-1}
\end{bmatrix}=\mathbf{F}\mathbf{A},
\end{equation}
$\mathbf{F}$ is the $N$-point DFT matrix with its $(l+1)$th row and $(n+1)$th column element $e^{-j2\pi\frac{l}{N}n}$ for $l,n=0,1,\ldots,N-1$, $\mathbf{G}_{k,m}=[{G}_{k,m}(0), {G}_{k,m}(1), \ldots, {G}_{k,m}(N-1)]^{\rm T}$,  $G_{k,m}(l)$ is the $N$-point DFT of $g_{k,m}(n)$, 
\begin{equation}
\label{Eqn23}
\mathbf{\Phi}={\rm diag}\left(e^{j\varphi 0} \; e^{j\varphi 1} \; \cdots  \; e^{j\varphi (N-1)}\right),
\end{equation} 
and $\varphi=-2\pi N_{\rm cp}/N$.

From \eqref{Eqn14}, \eqref{Eqn15}, and \eqref{Eqn18}, $\mathbf{b}_i$ can be calculated by
\begin{equation}
\label{Eqn24}
 \mathbf{b}_i= \mathbf{P}^{-1}_f \left( \mathbf{P}_1{\mathbf{d}}_{i-1}+\mathbf{P}_1\mathbf{A}^{-1}\mathbf{w}_{i-1}-\mathbf{P}_2\mathbf{d}_i \right).
\end{equation} 
 
\begin{figure} [!h]
     \centering
     %%----start of first subfigure----
     \subfloat[]{
   \label{fig:fig2a}        %% label for first subfigure
          \includegraphics[width=3in]{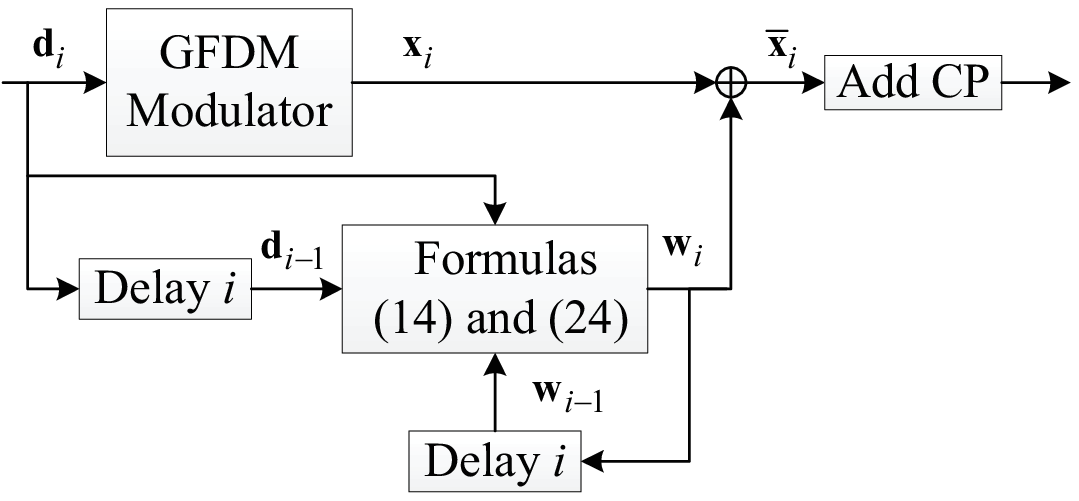}}
     \hspace{0.1\linewidth}  \\[20pt]  
     \subfloat[]{
          \label{fig:fig2b}        %% label for second subfigure
          \includegraphics[width=4in]{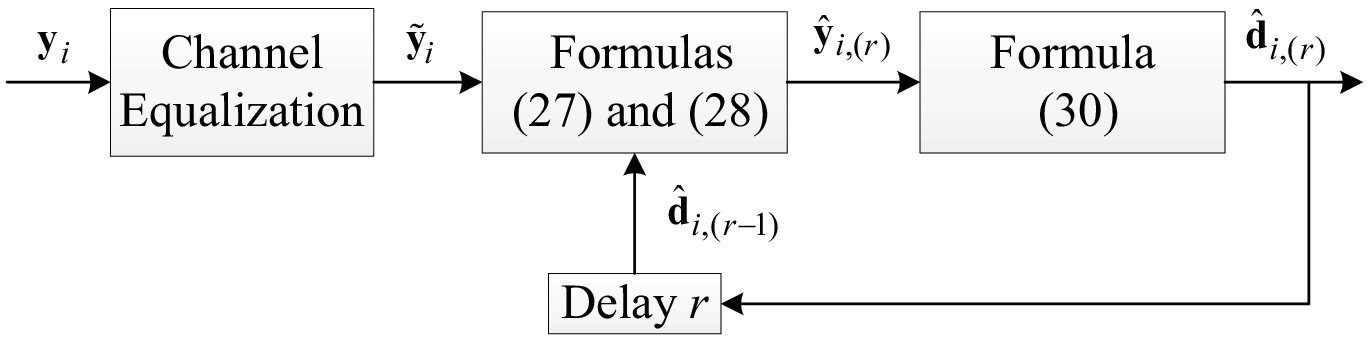}}  
 \caption{Block diagram of the TD-NC-GFDM transmitter and receiver: (a) Transmitter; (b) Receiver.}
     \label{fig:fig2}               %% label for entire figure
\end{figure}

Therefore, as shown in Fig. \ref{fig:fig2a}, the smoothed GFDM signal is expressed by
 \begin{equation}
\label{Eqn25}
 \bar{\mathbf{x}}_i= {\mathbf{x}}_i+ \mathbf{Q}\mathbf{P}^{-1}_f \left( \mathbf{P}_1{\mathbf{d}}_{i-1}+\mathbf{P}_1\mathbf{A}^{-1}\mathbf{w}_{i-1}-\mathbf{P}_2\mathbf{d}_i \right).
\end{equation}

At the receiver side, an iterative signal recovery algorithm is proposed to eliminate the interference caused by the smooth signal as shown in Fig. \ref{fig:fig2b}.

With the same assumption as \eqref{Eqn7}, after removing CP and equalizing the channel effect by ZF, the received TD-NC-GFDM signal can be expressed as 
 \begin{equation}
\label{Eqn26}
 \tilde{\mathbf{y}}_i= \bar{\mathbf{x}}_i+ \mathbf{H}^{-1}_i  \mathbf{n}_i.
\end{equation} 
  
 Our goal is to reconstruct the smooth signal and cancel it from the received TD-NC-GFDM signal. Suppose that the ZF GFDM receiver in \eqref{Eqn8} is employed. The estimated data $ \hat{\mathbf{y}}_{i, (r)}$  in the $r$th iteration of the signal recovery algorithm can be formulated  as  
  \begin{equation}
\label{Eqn27}
 \hat{\mathbf{y}}_{i, (r)}= \mathbf{A}^{-1} \left( \tilde{\mathbf{y}}_i - \mathbf{w}_{i, (r)} \right) ,
\end{equation} 
where the reconstructed smooth signal is
 \begin{equation}
\label{Eqn28}
\mathbf{w}_{i, (r)}= \mathbf{P}_{\rm w} \left( \mathbf{A}^{-1}\tilde{\mathbf{y}}_i \right) - \mathbf{Q}\mathbf{P}^{-1}_f \mathbf{P}_2 \hat{\mathbf{d}}_{i, (r-1)},
\end{equation} 
and the decoding matrix $\mathbf{P}_{\rm w}$ is 
 \begin{equation}
\label{Eqn29}
\mathbf{P}_{\rm w} = \mathbf{Q}\mathbf{P}^{-1}_f \mathbf{P}_2 ,
\end{equation} 
which is proved in Appendix A. $\hat{\mathbf{d}}_{i, (r-1)}$ is initialized as a all-zero vector and then is detected by the hard decision in constellation as
 \begin{equation}
\label{Eqn30}
\hat{d}_{i, k,m,(r)} =\arg\min\limits_{d\in\mathcal{C}}\left\{\left|  \hat{{y}}_{i, k,m,(r)} - d \right|^2 \right\},
\end{equation} 
where $\mathcal{C}$ denotes the constellation set.
 
%  Fig. \ref{fig:fig2} shows the TD-NC-GFDM transmitter and receiver.

%===================================================================================================================================================================
\subsection{Low-Interference TD-NC-GFDM}

\begin{figure} [!t]
     \centering
     %%----start of first subfigure----
     \subfloat[]{
   \label{fig:fig3a}        %% label for first subfigure
          \includegraphics[width=3.5in]{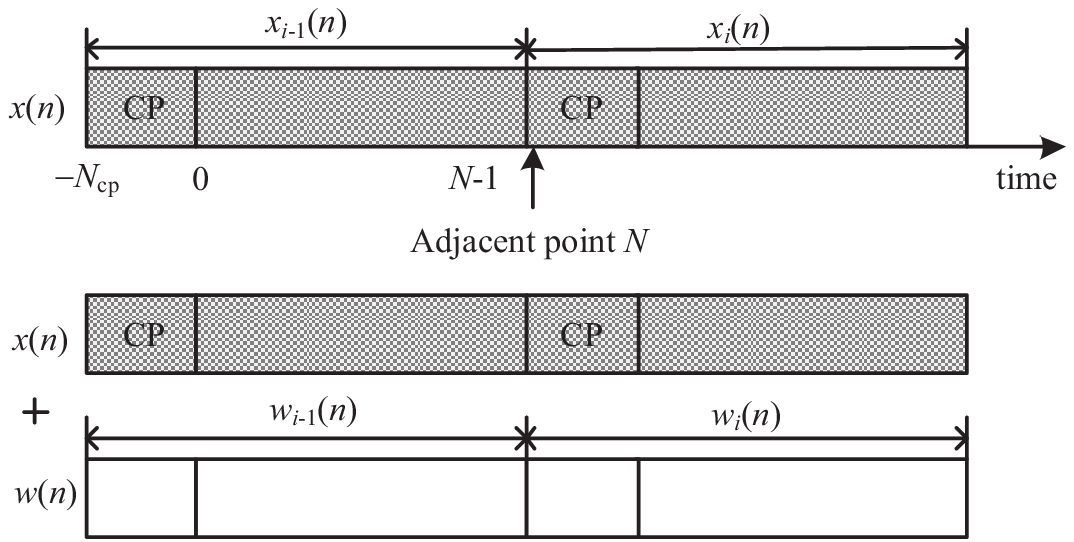}}
     \hspace{0.1\linewidth}  \\[20pt]  
     \subfloat[]{
          \label{fig:fig3b}        %% label for second subfigure
          \includegraphics[width=3.5in]{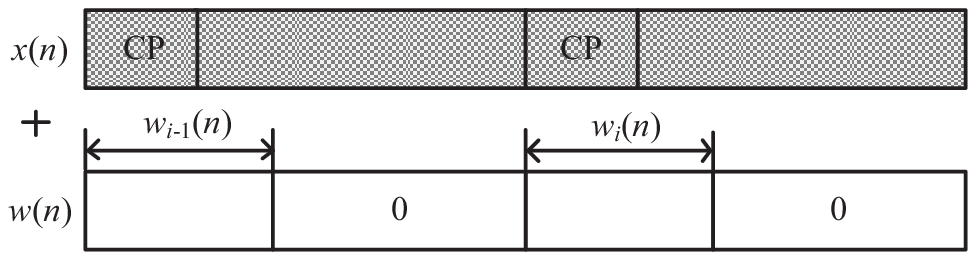}}  
\caption{Two different methods of adding the smooth signal to the GFDM signal in the time domain: (a) Addition in the whole GFDM symbol; (b) Addition in the front of each GFDM symbol.}
     \label{fig:fig3}               %% label for entire figure
\end{figure}

When the number of subcarrier $K$ or the number of GFDM subsymbol $M$ is increased, the complexity of the signal recovery algorithm for TD-NC-GFDM is high. Moreover, as shown in Fig. \ref{fig:fig3a}, the smooth signal in TD-NC-GFDM is distributed in the whole GFDM symbol, which causes high interference. Thus, to avoid the complexity of the signal recovery algorithm and reduce the interference caused by the smooth signal, as shown in Fig. \ref{fig:fig3b}, the smooth signal in the proposed low-interference TD-NC-GFDM is truncated in the time domain, aligned with the beginning of the CP-inserted GFDM symbol, and added in the front part of the CP-inserted GFDM symbol. In this case, to make the GFDM signal \emph{N}-continuous, $w_i(n)$ should satisfy 
\begin{equation}
  \label{Eqn31}
 \left.w^{(v)}_i(n)\right|_{n=-N_{\rm cp}}=\left.{x}^{(v)}_{i-1}(n)\right|_{n=N} - \left.x^{(v)}_{i}(n)\right|_{n=-N_{\rm cp}}.
\end{equation}

According to the addition of the smooth signal $w_i(n)$ in the low-interference TD-NC-GFDM, the linear-combination design of $w_i(n)$ is described as
\begin{equation}
  \label{Eqn32}
  w_i(n)=\left\{\begin{matrix}
          \sum\limits^{V}_{v=0}{{b}_{i,v}\tilde{f}_v(n)}, & n\in \mathcal{L}, \\
          0, & n\in \mathcal{N}\backslash \mathcal{L},
\end{matrix}\right.
\end{equation}
where $\mathcal{L}\triangleq\left\{-N_{\rm cp},-N_{\rm cp}+1,\ldots,-N_{\rm cp}+L-1\right\}$ indicates the location of $w_i(n)$ with length \emph{L}, and the basis signals $\tilde{f}_v(n)$ belong to the basis set $\tilde{\mathcal{Q}}$, defined as
\begin{align}
  \label{Eqn33}
  \tilde{\mathcal{Q}} \triangleq \left\{\tilde{\mathbf{q}}_{\tilde{v}}\left|\tilde{\mathbf{q}}_{\tilde{v}}
  =\left[\tilde{f}_{\tilde{v}}(-N_{\rm cp}),\tilde{f}_{\tilde{v}}(-N_{\rm cp}+1),\ldots, \right.   \tilde{f}_{\tilde{v}}(-N_{\rm cp}+L-1)\right]^{\rm T}, \tilde{v}\in \mathcal{U}_{2V}\right\}.
  \end{align}

On the one hand, the time duration of $\tilde{f}_{\tilde{v}}(n)$ is truncated by a preset window function $z(n)$, which is considered as a smooth and zero-edged window function, such as triangular, Hanning, or Blackman window function. Then, the truncated basis signals are given by
\begin{equation}
  \label{Eqn34}
\tilde{f}_{\tilde{v}}(n)=\left\{\begin{matrix}
          f_{\tilde{v}}(n) z(n) u(n), & n\in\mathcal{L},  \\
          0, & n\in\mathcal{N}\backslash \mathcal{L},
\end{matrix}\right.
\end{equation}
where $u(n)$ denotes the unit-step function and $z(n)$ is the right half of the baseband-equivalent window function.

On the other hand, by substituting \eqref{Eqn9} and \eqref{Eqn32} into \eqref{Eqn31}, the coefficients ${b}_{i,v}$ can be calculated as
\begin{equation}
  \label{Eqn35}
 \mathbf{b}_i= \mathbf{P}^{-1}_f \left( \mathbf{P}_1{\mathbf{d}}_{i-1} - \mathbf{P}_2\mathbf{d}_i \right).
\end{equation}

\begin{figure}[h]
\begin{center}
\includegraphics[width=3in]{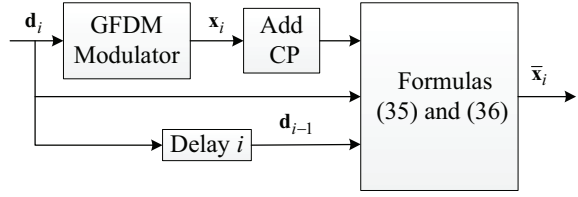}  
\caption{Block diagram of the low-interference TD-NC-GFDM transmitter.}
\label{fig:fig4}
\end{center}
\end{figure}

Finally, as shown in Fig. \ref{fig:fig3b}, $w_i(n)$ is superposed onto the front part of the CP-inserted GFDM symbol to achieve the \emph{N}-continuous GFDM signal $\bar{\mathbf{x}}_i$, given as
\begin{equation}
  \bar{\mathbf{x}}_i=\left\{\begin{matrix}
      \mathbf{x}_i+\left[\begin{array}{c}\tilde{\mathbf{Q}} {\mathbf{b}}_i \\ \mathbf{0}_{(N+N_{\rm cp}-L)\times 1}\end{array}\right], & 0\leq i\leq N_{\rm s}-1, \\
     \tilde{ \mathbf{Q}} {\mathbf{b}}_i, &  i=N_{\rm s},
\end{matrix}\right.
  \label{Eqn36}
\end{equation}
where $\tilde{\mathbf{Q}} =[\tilde{\mathbf{q}}_0 \; \tilde{\mathbf{q}}_1 \; \ldots \; \tilde{\mathbf{q}}_V]$, $N_{\rm s}$ is the number of the GFDM symbols, and $\mathbf{d}_{-1}$ is initialized as $\mathbf{d}_{-1}=\mathbf{0}_{N\times 1}$ under the assumption of the back edge of $\mathbf{x}_{-1}$ equal to zero. According to \eqref{Eqn35} and \eqref{Eqn36}, Fig. \ref{fig:fig4} depicts the low-interference TD-NC-GFDM transmitter.

%===================================================================================================================================================================
\subsection{SIR Analysis in TD-NC-GFDM}

Since the data $d_{i,k,m}$ follows i.i.d. process, we have $E\{\mathbf{d}_i \mathbf{d}^{\rm H}_i \}=\mathbf{I}_N$ and $E\{\mathbf{d}_i \mathbf{d}^{\rm H}_{i-1} \}=\mathbf{0}_{N\times N}$.  Firstly, we let
\begin{equation}
  \label{Eqn37}
\tilde{\mathbf{P}}= \mathbf{A}^{-1}\mathbf{Q}\mathbf{P}^{-1}_f \mathbf{P}_2=\mathbf{A}^{-1}\mathbf{P}_{\rm w} ,
\end{equation}
which has been proved to be an idempotent and Hermitian matrix in Appendix B.

\begin{figure}[h]
\begin{center}
\includegraphics[width=5in]{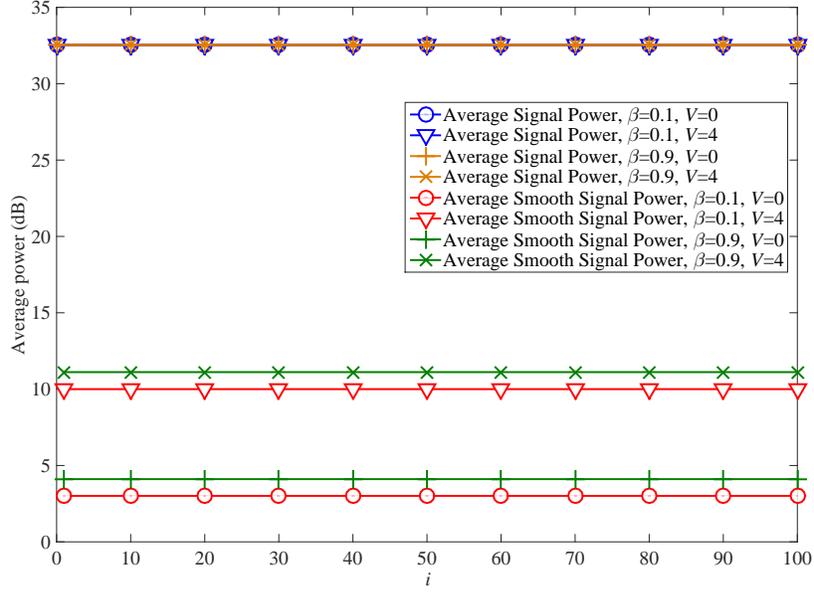}  
\caption{Average signal power $ E\left\{\bar{\mathbf{d}}_i \bar{\mathbf{d}}^{\rm H}_i \right\}$ and average smooth signal power $E\left\{ \left( \mathbf{A}^{-1}\mathbf{w}_i \right)^{\rm H}  \mathbf{A}^{-1}\mathbf{w}_i \right\}$ with different roll-off factors and HDOs, where the simulation parameters are given in Section IV.}
\label{fig:fig5}
\end{center}
\end{figure}

According to \eqref{Eqn25} and \eqref{Eqn37}, we have
\begin{align}
  \label{Eqn38}
E\left\{\bar{\mathbf{d}}_i \bar{\mathbf{d}}^{\rm H}_i \right\}&= E\left\{ \mathbf{A}^{-1} \bar{\mathbf{x}}_i \left( \mathbf{A}^{-1} \bar{\mathbf{x}}_i \right)^{\rm H} \right\}  \nonumber \\
&=E\left\{ \left( \left(\mathbf{I}_N - \tilde{\mathbf{P}} \right)\mathbf{d}_i +\mathbf{A}^{-1}\mathbf{Q}\mathbf{P}^{-1}_f \mathbf{P}_{1} \bar{\mathbf{d}}_{i-1} \right)  \left( \left(\mathbf{I}_N - \tilde{\mathbf{P}} \right)\mathbf{d}_i +\mathbf{A}^{-1}\mathbf{Q}\mathbf{P}^{-1}_f \mathbf{P}_{1} \bar{\mathbf{d}}_{i-1} \right)^{\rm H} \right\}  \nonumber \\
&=E\left\{  \left(\mathbf{I}_N - \tilde{\mathbf{P}} \right)\mathbf{d}_i   \mathbf{d}^{\rm H}_i \left(\mathbf{I}_N - \tilde{\mathbf{P}} \right)^{\rm H} \right\}  + E\left\{ \mathbf{A}^{-1}\mathbf{Q}\mathbf{P}^{-1}_f \mathbf{P}_{1} \bar{\mathbf{d}}_{i-1}\bar{\mathbf{d}}^{\rm H}_{i-1}   \left( \mathbf{A}^{-1}\mathbf{Q}\mathbf{P}^{-1}_f \mathbf{P}_{1}  \right)^{\rm H} \right\}  \nonumber \\
&=\mathbf{I}_N - \tilde{\mathbf{P}} + \mathbf{A}^{-1}\mathbf{Q}\mathbf{P}^{-1}_f \mathbf{P}_{1} E\left\{ \bar{\mathbf{d}}_{i-1}\bar{\mathbf{d}}^{\rm H}_{i-1}\right\} \mathbf{P}^{\rm H}_{1}  \left(\mathbf{P}^{-1}_f\right)^{\rm H} \mathbf{Q}^{\rm H}  \left( \mathbf{A}^{-1}  \right)^{\rm H}  \nonumber \\
&=\mathbf{I}_N - \tilde{\mathbf{P}} + \hat{\mathbf{P}}  E\left\{ \bar{\mathbf{d}}_{i-1}\bar{\mathbf{d}}^{\rm H}_{i-1}\right\} \hat{\mathbf{P}} ^{\rm H} ,
\end{align}
where 
\begin{equation}
  \label{Eqn39}
  \hat{\mathbf{P}}=\mathbf{A}^{-1}\mathbf{Q}\mathbf{P}^{-1}_f \mathbf{P}_{1}.
\end{equation}
Moreover, with the initialization of $\bar{\mathbf{d}}_{0}=\mathbf{d}_{0}$, it is inferred that  $E\{\bar{\mathbf{d}}_0 \bar{\mathbf{d}}^{\rm H}_0 \}=\mathbf{I}_N$. Thus, Eq. \eqref{Eqn38} can be further expressed by
\begin{align}
  \label{Eqn40}
E\left\{\bar{\mathbf{d}}_i \bar{\mathbf{d}}^{\rm H}_i \right\} = \sum\limits^{i}_{i_1=0}{\hat{\mathbf{P}}^{i_1} \left(\hat{\mathbf{P}}^{\rm H}\right)^{i_1}} - \sum\limits^{i-1}_{i_1=0}{\hat{\mathbf{P}}^{i_1}\tilde{\mathbf{P}} \left(\hat{\mathbf{P}}^{\rm H}\right)^{i_1}}
\end{align}
for $i\in\{0,1,\ldots, N_{\rm s} \}$. When the roll-off factor of $g(n)$ equals to zero, i.e., $\beta=0$,  it is proved in Appendix C that the elements $\bar{d}_{i,k,m}$ in $\bar{\mathbf{d}}_{i}$ are also uncorrelated. Fig. \ref{fig:fig5} shows that with different roll-off factors and HDOs, the average signal power  $E\left\{ \bar{\mathbf{d}}^{\rm H}_i \bar{\mathbf{d}}_i  \right\}={\rm Tr} \left\{ E\left\{\bar{\mathbf{d}}_i \bar{\mathbf{d}}^{\rm H}_i \right\} \right\}$ varies tiny in $i$. Thus, for different roll-off factors and HDOs, the average signal power can be approximated to be constant.

The average power of the smooth signal $\mathbf{A}^{-1}\mathbf{w}_i$ can be expressed by 
\begin{align}
  \label{Eqn41}
E\left\{ \left( \mathbf{A}^{-1}\mathbf{w}_i \right)^{\rm H}  \mathbf{A}^{-1}\mathbf{w}_i \right\}
&={\rm Tr}\left\{ E\left\{  \mathbf{A}^{-1}\mathbf{w}_i  \left(\mathbf{A}^{-1}\mathbf{w}_i \right)^{\rm H}  \right\} \right\}  \nonumber \\
&={\rm Tr}\left\{ E\left\{  \mathbf{A}^{-1}\mathbf{Q}\mathbf{P}^{-1}_{f} \left(\mathbf{P}_1\bar{\mathbf{d}}_{i-1}-\mathbf{P}_2\mathbf{d}_i \right) \left(\mathbf{P}_1\bar{\mathbf{d}}_{i-1}-\mathbf{P}_2\mathbf{d}_i \right)^{\rm H} \left( \mathbf{A}^{-1}\mathbf{Q}\mathbf{P}^{-1}_{f} \right)^{\rm H} \right\} \right\}  \nonumber \\
&={\rm Tr}\left\{  \hat{\mathbf{P}} E\left\{ \bar{\mathbf{d}}_{i-1}\bar{\mathbf{d}}^{\rm H}_{i-1}\right\}\hat{\mathbf{P}}^{\rm H} + \tilde{\mathbf{P}}  \tilde{\mathbf{P}} ^{\rm H}  \right\}.
\end{align}
According to \eqref{Eqn40}, Eq. \eqref{Eqn41} can be further written as 
\begin{align}
  \label{Eqn42}
E\left\{ \left( \mathbf{A}^{-1}\mathbf{w}_i \right)^{\rm H}  \mathbf{A}^{-1}\mathbf{w}_i \right\}
&={\rm Tr}\left\{  \hat{\mathbf{P}} E\left\{ \bar{\mathbf{d}}_{i-1}\bar{\mathbf{d}}^{\rm H}_{i-1}\right\}\hat{\mathbf{P}}^{\rm H} + \tilde{\mathbf{P}}  \tilde{\mathbf{P}} ^{\rm H}  \right\} \nonumber \\
&={\rm Tr}\left\{   \sum\limits^{i-1}_{i_1=0}{\hat{\mathbf{P}}^{i_1+1} \left(\hat{\mathbf{P}}^{\rm H}\right)^{i_1+1}} - \sum\limits^{i-2}_{i_1=0}{\hat{\mathbf{P}}^{i_1+1}\tilde{\mathbf{P}} \left(\hat{\mathbf{P}}^{\rm H}\right)^{i_1+1}} + \tilde{\mathbf{P}}  \tilde{\mathbf{P}} ^{\rm H}  \right\}
\end{align}
for $i\in\{1,2,\ldots, N_{\rm s}\}$. The smooth signal $\mathbf{w}_i$ is initialized by $\mathbf{w}_0=\mathbf{0}_{N\times 1}$. Additionally, when $\beta=0$, according to the proved relationship between $\hat{\mathbf{P}}$ and $\tilde{\mathbf{P}}$ in \eqref{Eqn63} proved in Appendix C and the property of idempotent matrix in \cite{Ref23}, we have
\begin{equation}
  \label{Eqn43}
E\left\{ \left( \mathbf{A}^{-1}\mathbf{w}_i \right)^{\rm H}  \mathbf{A}^{-1}\mathbf{w}_i \right\}
={\rm Tr}\left\{  \hat{\mathbf{P}}\hat{\mathbf{P}}^{\rm H} + \tilde{\mathbf{P}}  \tilde{\mathbf{P}} ^{\rm H}  \right\}
=2 \; {\rm Tr}\left\{   \tilde{\mathbf{P}} \right\}
=2 \; {\rm rank}\left\{ \tilde{\mathbf{P}}  \right\}
=2(V+1).
\end{equation}
As shown in Fig. \ref{fig:fig5}, for a big roll-off factor or HDO, the average smooth signal power in \eqref{Eqn42}  is increased, where HDO is the dominant factor. On the contrary, for a small roll-off factor or HDO, the average smooth signal power in \eqref{Eqn42} is reduced. Moreover, for the reduced roll-off factor, the average smooth signal power approaches to $2(V+1)$ in \eqref{Eqn43}. 

\begin{figure}[!h]
\begin{center}
\includegraphics[width=5in]{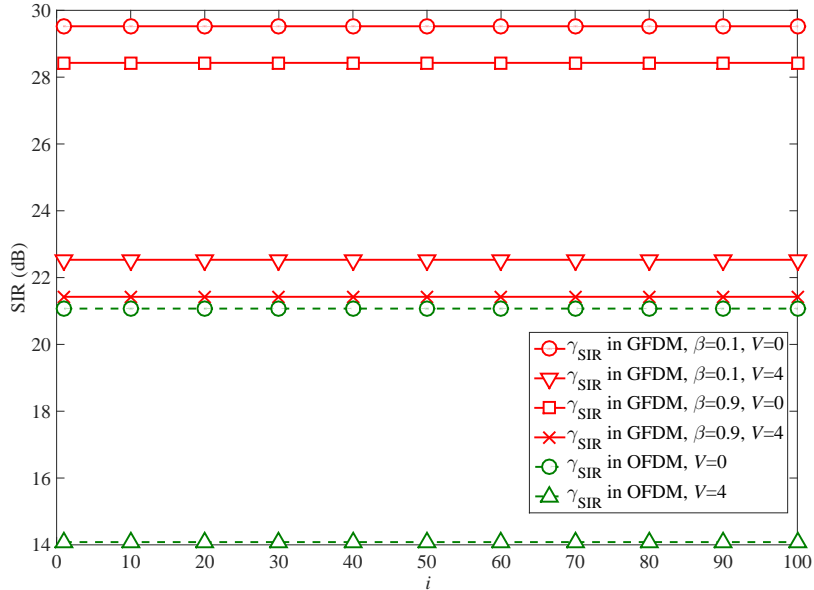}  
\caption{SIRs of the TD-NC-GFDM signal and TD-NC-OFDM signal with different roll-off factors and HDOs, where the simulation parameters are given in Section IV.}
\label{fig:fig6}
\end{center}
\end{figure}

Thus, the SIR between the original GFDM data $\mathbf{d}_{i}$ and the smooth signal $\mathbf{A}^{-1}\mathbf{w}_i$ is calculated by
\begin{align}
  \label{Eqn44}
\gamma_{\rm SIR}
%&=\frac{E\left\{ \mathbf{d}_{i}^{\rm H}  \mathbf{d}_{i} \right\}} {E\left\{ \left( \mathbf{A}^{-1}\mathbf{w}_i \right)^{\rm H}  \mathbf{A}^{-1}\mathbf{w}_i \right\}}  \nonumber \\
&=\frac{{\rm Tr} \left\{ E\left\{ \mathbf{d}_{i} \mathbf{d}_{i}^{\rm H} \right\}\right\}} {E\left\{ \left( \mathbf{A}^{-1}\mathbf{w}_i \right)^{\rm H}  \mathbf{A}^{-1}\mathbf{w}_i \right\}}  \nonumber \\
&=\frac{N} { {\rm Tr}\left\{ \sum\limits^{i-1}_{i_1=0}{\hat{\mathbf{P}}^{i_1+1} \left(\hat{\mathbf{P}}^{\rm H}\right)^{i_1+1}} - \sum\limits^{i-2}_{i_1=0}{\hat{\mathbf{P}}^{i_1+1}\tilde{\mathbf{P}} \left(\hat{\mathbf{P}}^{\rm H}\right)^{i_1+1}} + \tilde{\mathbf{P}}  \tilde{\mathbf{P}} ^{\rm H}  \right\} }. 
\end{align}
When $\beta=0$, based on \eqref{Eqn43}, $\gamma_{\rm SIR}$ is expressed as 
\begin{equation}
  \label{Eqn45}
\gamma_{\rm SIR}=\frac{N} { 2(V+1) }. 
\end{equation}
Eq. \eqref{Eqn45} has $M$ times larger than the SIR $\frac{K}{2(V+1)}$ in TD-NC-OFDM \cite{Ref22}. Fig. \ref{fig:fig6} compares the SIRs of the TD-NC-GFDM signal and TD-NC-OFDM signal with varying roll-off factors and HODs. For a small HDO or roll-off factor, the SIR of TD-NC-GFDM is increased, where HDO has the dominant effect on SIR. On the contrary, as the HDO or roll-off factor is increased, the SIR of TD-NC-GFDM is degraded. Moreover, for the same data rate, compared to $M$ many $K$-point TD-NC-OFDM signal, the SIR of the $MK$-point TD-NC-GFDM signal is higher than that of the TD-NC-OFDM signal. For the small roll-off factor, the enhanced SIR gain of TD-NC-GFDM is almost up to $M$ times compared to TD-NC-OFDM.

%%%%%%%%%%%%%%%%%%%%%%%%%%%%%%%%%%%%%%%%%%%%%%%%%%%%%%%%%%%%%%%%%%%%%%%%%%%%%%%%%%%%%%%%%%%%%%%%%%%%%%%%%%%%%%%%
\section{Simulation Results}

In the following simulations, the system parameters are listed in Table \ref{table1}. The 9-path EVA channel model in 3GPP LTE is used, whose channel delay and channel power are [0, 30, 150, 310, 370, 710, 1090, 1730, 2510] ns and [0, -1.5, -1.4, -3.6, -0.6, -9.1, -7.0, -12.0, -16.9] dB, respectively.  To avoid the edge of the window in the windowing technique for GFDM \cite{Ref1} exceeding the given lengths of CP and cyclic postfix, in the current system parameters, the roll-off factor of the window in the windowing technique is set to be 0.015.

\renewcommand\arraystretch{1}
\begin{table} [h]
  \centering 
  \caption{Simulation Parameters }
  \label{table1}
  \begin{tabular}{|c|c|}
\hline
% after \\ : \hline or \cline{col1-col2} \cline{col3-col4} ...
\textbf{Parameters} & \textbf{Values} \\
 \hline
 Constellation modulation  & 16QAM \\
  Transmitter filter & RC \\
  Roll-off factor ($\beta$) & 0.1 \\%and 0.9    \\
  Number of subcarriers (\emph{K}) & 256    \\
  Number of subsymbols (\emph{M}) & 7    \\
 Time-domain oversampling factor  & 4 \\
  Frequency interval ($\Delta f$) & 15 KHz \\
  Sampling interval & 9.3 ns \\%0.1488 $\mu$s \\
  Carrier frequency & 2 GHz \\
%Channel code & Convolutional code \\
%Code rate & 0.5 \\
  Maximum Doppler shift ($f_D$) & 100 Hz \\
  Length of CP in GFDM &  280\\
  Length of CP in OFDM & 280  \\
  Channel environment & Multipath Rayleigh fading channel \\
\hline
\end{tabular}
\end{table}

\begin{figure}[!h]
\begin{center}
\includegraphics[width=5in]{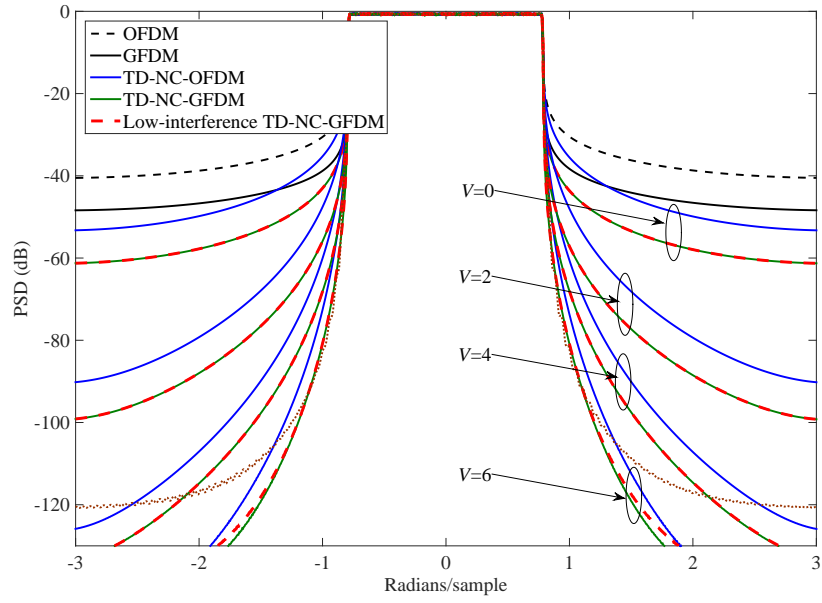}  
\caption{PSD comparison among TD-NC-OFDM, the windowing technique, TD-NC-GFDM, and the low-interference TD-NC-GFDM, where $L=280$ equal to the CP length in the low-interference TD-NC-GFDM, and the lengths of CP and cyclic postfix of the windowing technique are both 140.}
\label{fig:fig7}
\end{center}
\end{figure}

Fig. \ref{fig:fig7} compares the PSDs among TD-NC-OFDM, the windowing technique for GFDM \cite{Ref1}, and TD-NC-GFDM and its low-interference scheme with different HDOs. With the HDO increasing, the sidelobe suppression performance of TD-NC-GFDM is significantly improved. For example, when $V=6$, TD-NC-GFDM and its low-interference scheme has more rapid sidelobe decaying than the windowing technique. It is also shown that TD-NC-GFDM and its low-interference scheme have much better sidelobe suppression performance than TD-NC-OFDM, since the GFDM transmitter filter has much lower sidelobes than the rectangularly pulsed OFDM waveform.

\begin{figure}[!h]
\begin{center}
\includegraphics[width=5in]{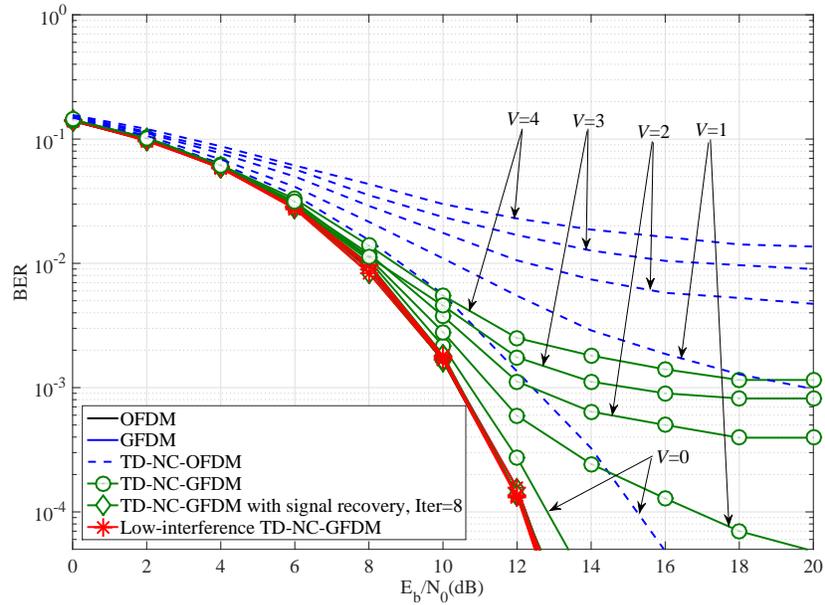}  
\caption{BERs of TD-NC-OFDM, TD-NC-GFDM, and the low-interference TD-NC-GFDM in AWGN channel.}
\label{fig:fig8}
\end{center}
\end{figure}

\begin{figure}[!h]
\begin{center}
\includegraphics[width=5in]{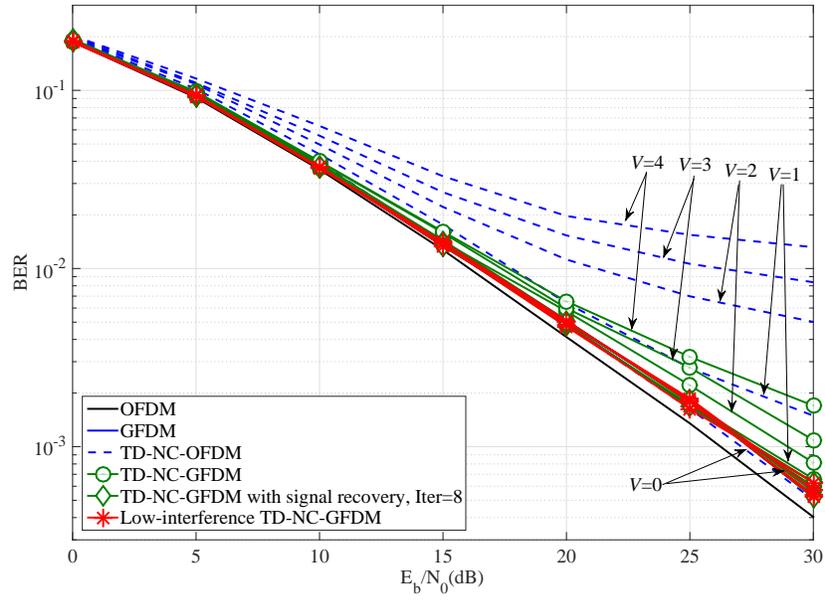}  
\caption{BERs of TD-NC-OFDM, TD-NC-GFDM, and the low-interference TD-NC-GFDM in Rayleigh fading channel.}
\label{fig:fig9}
\end{center}
\end{figure}

\begin{figure}[!h]
\begin{center}
\includegraphics[width=5in]{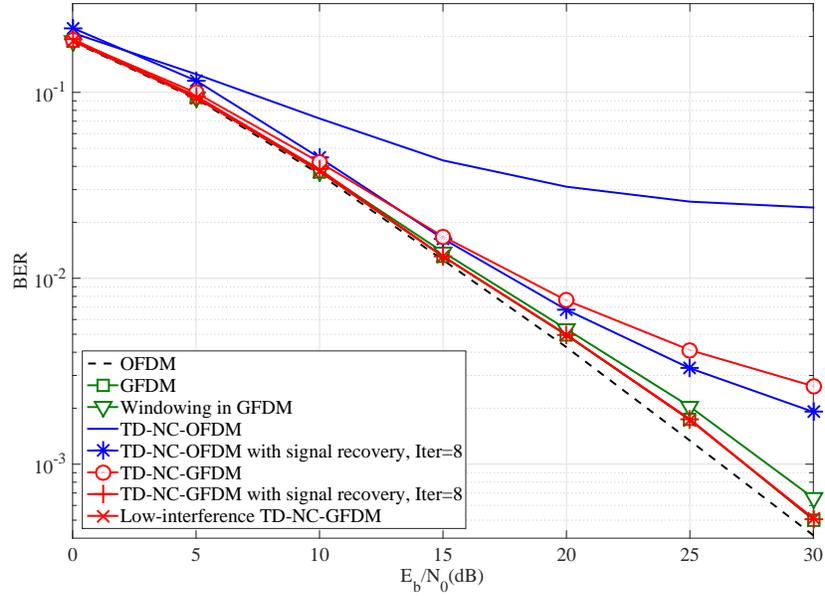}  
\caption{BERs of TD-NC-OFDM, the windowing technique, TD-NC-GFDM, and the low-interference TD-NC-GFDM in Rayleigh fading channel, where $V=6$.}
\label{fig:fig10}
\end{center}
\end{figure}

Figs. \ref{fig:fig8} and \ref{fig:fig9} compare the BER performance of TD-NC-OFDM, TD-NC-GFDM, and the low-interference TD-NC-GFDM in AWGN channel and Rayleigh fading channel, respectively. Matching the SIR analysis of the TD-NC-GFDM signal in Section III-C, the BER performance of TD-NC-GFDM is degraded by the increased HDO and TD-NC-GFDM has much better BER performance than TD-NC-OFDM. After the iterative signal recovery for the received TD-NC-GFDM signal, the BER performance is significantly improved as good as the original GFDM. Moreover, without the signal recovery, the low-interference TD-NC-GFDM can obtain the same BER as the original GFDM signal, while maintaining as good sidelobe suppression performance as TD-NC-GFDM as shown in Fig. \ref{fig:fig7}. On the other hand, Fig. \ref{fig:fig10} further compares the BERs among TD-NC-OFDM, the windowing technique, TD-NC-GFDM and the low-interference TD-NC-GFDM in Rayleigh fading channel, where a big HDO $V=6$ is considered. In this case, the signal recovery of TD-NC-OFDM \cite{Ref22} cannot completely reduce the high interference caused by the smooth signal in the limited number of iterations. However, for the higher SIR in TD-NC-GFDM, the data can be well recovered by the signal recovery algorithm in Section III-A. Meanwhile, the low-interference TD-NC-GFDM can also achieve the same BER performance as the original GFDM, whose $\rm E_b/N_0$ is 1dB better than the windowing technique when BER=$10^{-3}$. Additionally, GFDM with the small roll-off factor, such as $\beta=0.1$, has close BER performance to OFDM. Thus, the low-interference TD-NC-GFDM is a good choice of sidelobe suppression in the future communications.

%\vspace{-0.5cm}

%\vspace{-1cm}

%%%%%%%%%%%%%%%%%%%%%%%%%%%%%%%%%%%%%%%%%%%%%%%%%%%%%%%%%%%%%%%%%%%%%%%%%%%%%%%%%%%%%%%%%%%%%%%%%%%%%%%%%%%%%%%%
\section{Conclusion}

This paper proposed the TD-NC-GFDM technique to suppress spectral leakage by adding a smooth signal. Based on TD-NC-OFDM, the linear-combination format of the smooth signal was adopted. The basis signals for the smooth signal were designed according to the GFDM transmitter filters and their high-order derivatives. To eliminate the interference caused by the smooth signal, we first proposed the iterative signal recovery algorithm for the received TD-NC-GFDM signal by the decoding matrix. Then, to avoid the high complexity of the signal recovery algorithm, the low-interference TD-NC-GFDM was presented by truncating the smooth signal. Furthermore, we analyzed that TD-NC-GFDM slightly affect the average power of GFDM signal, and more importantly, the SIR of TD-NC-GFDM is higher than that of TD-NC-OFDM when the number of GFDM subsymbols is bigger than 1. Simulation results showed that with the small roll-off factor $\beta=0.1$, TD-NC-GFDM and the low-interference TD-NC-GFDM could achieve significant sidelobe suppression better than TD-NC-OFDM, and the low-interference TD-NC-GFDM can avoid the signal recovery complexity while maintaining the same BER performance as the original GFDM, which has the close BER to OFDM.

%%%%%%%%%%%%%%%%%%%%%%%%%%%%%%%%%%%%%%%%%%%%%%%%%%%%%%%%%%%%%%%%%%%%%%%%%%%%%%%%%%%%%%%%%%%%%%%%%%%%%%%%%%%%%%%%
\appendices

%===================================================================================================================================================================
\section{Proof of the Decoding Matrices in \eqref{Eqn29} }

\subsection{Necessary Condition}

According to the design of the smooth signal in \eqref{Eqn24} (or \eqref{Eqn17}), the $i$th smooth signal is related to the $(i-1)$th and $i$th GFDM symbols. When $r=0$, the initialization is $\hat{\mathbf{d}}_{i,(r)}=\mathbf{0}_{N\times 1}$. In this case, the $i$th GFDM symbol is known. To obtain the $(i-1)$th GFDM data, a matrix $\mathbf{P}_{\rm w}$ is used for the received TD-NC-GFDM signal in \eqref{Eqn26} as  
\begin{align}
  \label{Eqn46}
\mathbf{P}_{\rm w}\mathbf{A}^{-1}\tilde{\mathbf{y}}_i &= \mathbf{P}_{\rm w}\mathbf{A}^{-1} \left( \bar{\mathbf{x}}_i+ \mathbf{H}^{-1}_i  \mathbf{n}_i \right) \nonumber \\
&=\mathbf{P}_{\rm w} {\mathbf{d}}_i + \mathbf{P}_{\rm w}\mathbf{A}^{-1}{\mathbf{w}}_i + \mathbf{P}_{\rm w}\mathbf{A}^{-1}\mathbf{H}^{-1}_i  \mathbf{n}_i  \nonumber \\
&=\mathbf{P}_{\rm w} {\mathbf{d}}_i + \mathbf{P}_{\rm w}\mathbf{A}^{-1} \mathbf{Q}\mathbf{P}^{-1}_f \left( \mathbf{P}_1 \bar{\mathbf{d}}_{i-1} - \mathbf{P}_2\mathbf{d}_i \right) + \mathbf{P}_{\rm w}\mathbf{A}^{-1}\mathbf{H}^{-1}_i  \mathbf{n}_i.
\end{align}
To extract the $(i-1)$th data, the following relationship should be satisfied 
\begin{equation}
  \label{Eqn47}
\mathbf{P}_{\rm w} = \mathbf{P}_{\rm w}\mathbf{A}^{-1} \mathbf{Q}\mathbf{P}^{-1}_f \mathbf{P}_2.
\end{equation}
Then, we have
\begin{equation}
  \label{Eqn48}
\mathbf{P}_{\rm w}\mathbf{A}^{-1}\tilde{\mathbf{y}}_i = \mathbf{P}_{\rm w}\mathbf{A}^{-1} \mathbf{Q}\mathbf{P}^{-1}_f \mathbf{P}_1 \bar{\mathbf{d}}_{i-1}  + \mathbf{P}_{\rm w}\mathbf{A}^{-1}\mathbf{H}^{-1}_i  \mathbf{n}_i.
\end{equation}
Similar to \eqref{Eqn14}, \eqref{Eqn17}, and \eqref{Eqn24}, the smooth signal can be reconstructed as
\begin{align}
  \label{Eqn49}
\mathbf{A}^{-1} \mathbf{w}_{i, (r)} &= \mathbf{A}^{-1}\mathbf{P}_{\rm w}\mathbf{A}^{-1}\tilde{\mathbf{y}}_i -  \mathbf{A}^{-1} \mathbf{Q}\mathbf{P}^{-1}_f \mathbf{P}_2 \hat{\mathbf{d}}_{i,(r)} \nonumber \\
&= \mathbf{A}^{-1} \mathbf{P}_{\rm w}\mathbf{A}^{-1} \mathbf{Q}\mathbf{P}^{-1}_f \mathbf{P}_1 \bar{\mathbf{d}}_{i-1}  -  \mathbf{A}^{-1} \mathbf{Q}\mathbf{P}^{-1}_f \mathbf{P}_2 \hat{\mathbf{d}}_{i,(r)} + \mathbf{A}^{-1}\mathbf{P}_{\rm w}\mathbf{A}^{-1}\mathbf{H}^{-1}_i  \mathbf{n}_i .
\end{align}
To obtain the smooth signal from the first two terms in \eqref{Eqn49}, compared to \eqref{Eqn17} and \eqref{Eqn24}, the following relationship should be satisfied
\begin{equation}
  \label{Eqn50}
\mathbf{P}_{\rm w}\mathbf{A}^{-1} \mathbf{Q}\mathbf{P}^{-1}_f  = \mathbf{Q}\mathbf{P}^{-1}_f .
\end{equation}

According to \eqref{Eqn47} and \eqref{Eqn50}, we have 
\begin{equation}
  \label{Eqn51}
\mathbf{P}_{\rm w} = \left(\mathbf{P}_{\rm w}\mathbf{A}^{-1} \mathbf{Q}\mathbf{P}^{-1}_f \right) \mathbf{P}_2 =\mathbf{Q}\mathbf{P}^{-1}_f \mathbf{P}_2.
\end{equation}

\subsection{Sufficient Condition}

According to \eqref{Eqn13} and \eqref{Eqn20}-\eqref{Eqn23}, we can obtain
\begin{align}
  \label{Eqn52}
&\mathbf{P}_2 \mathbf{A}^{-1}\mathbf{Q}  \nonumber \\
&=\mathbf{B}\mathbf{\Phi}\mathbf{F}\mathbf{A}\mathbf{A}^{-1}\mathbf{Q} \nonumber \\
&=\mathbf{B}\mathbf{\Phi}(\mathbf{F}\mathbf{Q}) \nonumber \\
&=\mathbf{B}\mathbf{\Phi} 
\begin{bmatrix}
F_0(0) e^{-j\varphi 0}   &  j2\pi\frac{0}{N} F_0(0) e^{-j\varphi 0} &  \cdots   & \left( j2\pi\frac{0}{N} \right)^V F_0(0) e^{-j\varphi 0} \\
F_0(1) e^{-j\varphi 1}   &   j2\pi\frac{1}{N} F_0(1) e^{-j\varphi 1}  &  \cdots    & \left( j2\pi\frac{1}{N} \right)^V F_0(1) e^{-j\varphi 1}  \\
\vdots &  \vdots & {} & \vdots \\
F_0(N-1) e^{-j\varphi (N-1)}  & j2\pi\frac{N-1}{N} F_0(N-1) e^{-j\varphi (N-1)} &  \cdots  & \left( j2\pi\frac{N-1}{N} \right)^V F_0(N-1) e^{-j\varphi (N-1)} \\
\end{bmatrix} \nonumber \\
&=\frac{1}{N} \begin{bmatrix}
1  &  1 &  \cdots   & 1 \\
j2\pi \frac{0}{N}   &    j2\pi \frac{1}{N}  &  \cdots    &  j2\pi \frac{N-1}{N} \\
\vdots &  \vdots & {} & \vdots \\
\left( j2\pi \frac{0}{N} \right)^V  & \left( j2\pi \frac{1}{N} \right)^V  &  \cdots  &  \left( j2\pi \frac{N-1}{N} \right)^V \\
\end{bmatrix} %\nonumber \\
\begin{bmatrix}
F_0(0)  &  j2\pi\frac{0}{N} F_0(0) &  \cdots   & \left( j2\pi\frac{0}{N} \right)^V F_0(0) \\
F_0(1)  &   j2\pi\frac{1}{N} F_0(1)  &  \cdots    & \left( j2\pi\frac{1}{N} \right)^V F_0(1) \\
\vdots &  \vdots & {} & \vdots \\
F_0(N-1)  & j2\pi\frac{N-1}{N} F_0(N-1)  &  \cdots  & \left( j2\pi\frac{N-1}{N} \right)^V F_0(N-1)  \\
\end{bmatrix}  \nonumber \\
&= \begin{bmatrix}
\frac{1}{N}\sum\limits^{N-1}_{l=0}{F_0(l)}  & \frac{1}{N}\sum\limits^{N-1}_{l=0}{ j2\pi \frac{l}{N} F_0(l)} &  \cdots   & \frac{1}{N}\sum\limits^{N-1}_{l=0}{ \left(j2\pi \frac{l}{N}\right)^V F_0(l)} \\
\frac{1}{N}\sum\limits^{N-1}_{l=0}{ j2\pi \frac{l}{N} F_0(l)} &  \frac{1}{N}\sum\limits^{N-1}_{l=0}{ \left(j2\pi \frac{l}{N}\right)^2 F_0(l)}  &  \cdots    & \frac{1}{N}\sum\limits^{N-1}_{l=0}{ \left(j2\pi \frac{l}{N}\right)^{V+1} F_0(l)} \\
\vdots &  \vdots & {} & \vdots \\
\frac{1}{N}\sum\limits^{N-1}_{l=0}{ \left(j2\pi \frac{l}{N}\right)^V F_0(l)}  & \frac{1}{N}\sum\limits^{N-1}_{l=0}{ \left(j2\pi \frac{l}{N}\right)^{V+1} F_0(l)}  &  \cdots  & \frac{1}{N}\sum\limits^{N-1}_{l=0}{ \left(j2\pi \frac{l}{N}\right)^{2V} F_0(l)}  \\
\end{bmatrix}  \nonumber \\
&=\begin{bmatrix}
f(-N_{\rm cp}) & f_1(-N_{\rm cp}) & \cdots  & f_V(-N_{\rm cp})\\
f_1(-N_{\rm cp}) & f_2(-N_{\rm cp}) & \cdots \!\! & f_{V+1}(-N_{\rm cp})\\
\vdots & \vdots & {}& \vdots \\
f_V(-N_{\rm cp}) & f_{V+1}(-N_{\rm cp}) &  \cdots & f_{2V}(-N_{\rm cp})
\end{bmatrix}   \nonumber \\
&=\mathbf{P}_f.
\end{align}

According to \eqref{Eqn52}, it is easy to prove that the two conditions in \eqref{Eqn47} and \eqref{Eqn50} can be satisfied. Thus, the matrix $\mathbf{P}_{\rm w}$ is correct and can be used as the decoding matrix in the signal recovery algorithm in TD-NC-GFDM receiver. 

%===================================================================================================================================================================
\section{Properties of Matrix $\tilde{\mathbf{P}}$ in \eqref{Eqn37}}

From \eqref{Eqn52}, we can derive that
\begin{equation}
  \label{Eqn53}
\tilde{\mathbf{P}}^2= \mathbf{A}^{-1}\mathbf{Q}\mathbf{P}^{-1}_f \mathbf{P}_2 \mathbf{A}^{-1}\mathbf{Q}\mathbf{P}^{-1}_f \mathbf{P}_2=\mathbf{A}^{-1}\mathbf{Q}\mathbf{P}^{-1}_f \mathbf{P}_f \mathbf{P}^{-1}_f \mathbf{P}_2=\tilde{\mathbf{P}}, 
\end{equation}
which denotes that $\tilde{\mathbf{P}}$ is idempotent.

Based on \eqref{Eqn37} and \eqref{Eqn52}, we have
\begin{equation}
  \label{Eqn54}
\mathbf{P}_2 \tilde{\mathbf{P}}= \mathbf{P}_2.
\end{equation}
Thus, $\tilde{\mathbf{P}}$ can also be expressed as
\begin{equation}
  \label{Eqn55}
 \tilde{\mathbf{P}}=\mathbf{P}^{\dagger}_2 \mathbf{P}_2=\mathbf{P}^{\rm H}_2 \left( \mathbf{P}_2 \mathbf{P}^{\rm H}_2\right)^{-1} \mathbf{P}_2,
\end{equation}
where $\mathbf{P}^{\dagger}_2$ denotes the Moore-Penrose pseudoinverse of $\mathbf{P}_2$ \cite{Ref23}, $\mathbf{P}^{\dagger}_2=\mathbf{P}^{\rm H}_2 \left( \mathbf{P}_2 \mathbf{P}^{\rm H}_2\right)^{-1}$. Eq. \eqref{Eqn55} shows that $\tilde{\mathbf{P}}$ is also a Hermitian matrix, i.e.,
\begin{equation}
  \label{Eqn56}
\tilde{\mathbf{P}}^{\rm H}=\tilde{\mathbf{P}}.
\end{equation}

%===================================================================================================================================================================
\section{Proof of the Uncorrelated Elements in $\bar{\mathbf{d}}_{i,k,m}$ for $\beta=0$ }

When $\beta=0$, the prototype filter $g(n)$ becomes a sinc function. Thus, under the constraint of normalized energy of $g(n)$, i.e., $\sum\limits^{N-1}_{n=0}{|g(n)|^2}=1$, the transmitter matrix $\mathbf{A}$ can be expressed by
\begin{equation}
  \label{Eqn57}
{\mathbf{A}}=\frac{1}{N} \mathbf{F}^{\rm H} \begin{bmatrix}  \mathbf{\Phi}_0\mathbf{C} & \mathbf{\Phi}_1\mathbf{C} &\cdots & \mathbf{\Phi}_{M-1}\mathbf{C} \end{bmatrix},
\end{equation}
where $\mathbf{\Phi}_m={\rm diag}\left( e^{-j2\pi\frac{m}{M}0} \; e^{-j2\pi\frac{m}{M}1} \; \cdots \; e^{-j2\pi\frac{m}{M}(N-1)} \right)$, 
\begin{equation}
  \label{Eqn58}
\mathbf{C}=\begin{bmatrix} 
\mathbf{c} & {} & {} & {} \\
 {} & \mathbf{c} & {} & {} \\
{} & {} & \ddots & {} \\
{}& {} & {} &\mathbf{c} \\
 \end{bmatrix}_{N\times K},
 \end{equation}
 and $\mathbf{c}$ is an $M\times 1$ all-ones vector $\mathbf{c}=\sqrt{K}\;[1 \; 1 \; \cdots \; 1]^{\rm T}$.

%Then, we will prove that $\mathbf{A}^{\rm H}=\mathbf{A}^{-1}$. 
According to \eqref{Eqn57} and \eqref{Eqn58}, we can calculate that
\begin{align}
  \label{Eqn59}
\mathbf{A}^{\rm H}\mathbf{A}&=\frac{1}{N^2}  \begin{bmatrix}  \mathbf{C}^{\rm H}\mathbf{\Phi}^{\rm H}_0 \\ \mathbf{C}^{\rm H}\mathbf{\Phi}^{\rm H}_1 \\ \vdots \\ \mathbf{C}^{\rm H}\mathbf{\Phi}^{\rm H}_{M-1} \end{bmatrix}  \mathbf{F}
\mathbf{F}^{\rm H} \begin{bmatrix}  \mathbf{\Phi}_0\mathbf{C} & \mathbf{\Phi}_1\mathbf{C} &\cdots & \mathbf{\Phi}_{M-1}\mathbf{C} \end{bmatrix}  \nonumber \\
&=\frac{1}{N} \begin{bmatrix}  
\mathbf{C}^{\rm H}\mathbf{\Phi}^{\rm H}_0\mathbf{\Phi}_0\mathbf{C} & \mathbf{C}^{\rm H}\mathbf{\Phi}^{\rm H}_0 \mathbf{\Phi}_1\mathbf{C} & \cdots & \mathbf{C}^{\rm H}\mathbf{\Phi}^{\rm H}_0 \mathbf{\Phi}_{M-1}\mathbf{C}\\ 
\mathbf{C}^{\rm H}\mathbf{\Phi}^{\rm H}_1\mathbf{\Phi}_0\mathbf{C} & \mathbf{C}^{\rm H}\mathbf{\Phi}^{\rm H}_1 \mathbf{\Phi}_1\mathbf{C} & \cdots & \mathbf{C}^{\rm H}\mathbf{\Phi}^{\rm H}_1 \mathbf{\Phi}_{M-1}\mathbf{C}\\ 
 \vdots & \vdots & {} & \vdots \\ 
\mathbf{C}^{\rm H}\mathbf{\Phi}^{\rm H}_{M-1}\mathbf{\Phi}_0\mathbf{C} & \mathbf{C}^{\rm H}\mathbf{\Phi}^{\rm H}_{M-1} \mathbf{\Phi}_1\mathbf{C} & \cdots & \mathbf{C}^{\rm H}\mathbf{\Phi}^{\rm H}_{M-1} \mathbf{\Phi}_{M-1}\mathbf{C}\\ 
\end{bmatrix}.
\end{align}

In \eqref{Eqn59}, for the structure of $\mathbf{C}$ in \eqref{Eqn58}, $\mathbf{C}^{\rm H}\mathbf{\Phi}^{\rm H}_m \mathbf{\Phi}_{m^{\prime}}\mathbf{C}$ is a diagonal matrix with its $(k+1)$th row and $(k+1)$th column element given by 
\begin{equation}
  \label{Eqn60}
K \sum\limits^{(k+1)M-1}_{l=kM}{e^{j2\pi\frac{m-m^{\prime}}{M}}l}=K \frac{1-e^{j2\pi(m-m^{\prime})}}{1-e^{j2\pi\frac{m-m^{\prime}}{M}}}=\begin{cases}
   KM,   & \text{ $m=m^{\prime}$}, \\
    0,  & \text{$m\neq m^{\prime}$},
\end{cases}
\end{equation}
where $k=0,1,\ldots,K-1$ and $m,m^{\prime}=0,1,\ldots,M-1$. Thus, Eq. \eqref{Eqn59} can be rewritten by 
\begin{equation}
  \label{Eqn61}
\mathbf{A}^{\rm H}\mathbf{A}=\mathbf{I}_N,
\end{equation}
which denotes that when $\beta=0$, $\mathbf{A}^{\rm H}=\mathbf{A}^{-1}$.

Based on the mathematical induction, with the initialization of $E\left\{\bar{\mathbf{d}}_0 \bar{\mathbf{d}}^{\rm H}_0 \right\}=\mathbf{I}_N$, we assume that $E\left\{\bar{\mathbf{d}}_{i-1} \bar{\mathbf{d}}^{\rm H}_{i-1} \right\}=\mathbf{I}_N$. From \eqref{Eqn19}-\eqref{Eqn23}, \eqref{Eqn37}, \eqref{Eqn39} and \eqref{Eqn55}, Eq. \eqref{Eqn38} can be expressed by
\begin{align}
  \label{Eqn62}
E\left\{\bar{\mathbf{d}}_i \bar{\mathbf{d}}^{\rm H}_i \right\}
&= \mathbf{I}_N - \tilde{\mathbf{P}} + \hat{\mathbf{P}} \hat{\mathbf{P}} ^{\rm H} \nonumber  \\
&= \mathbf{I}_N - \mathbf{P}^{\rm H}_2 \left( \mathbf{P}_2 \mathbf{P}^{\rm H}_2\right)^{-1} \mathbf{P}_2 + \left( \mathbf{A}^{-1}\mathbf{Q}\mathbf{P}^{-1}_f \right) \mathbf{P}_{1} \mathbf{P}^{\rm H}_{1}  \left( \mathbf{A}^{-1}\mathbf{Q}\mathbf{P}^{-1}_f \right)^{\rm H} \nonumber \\
&= \mathbf{I}_N - \mathbf{P}^{\rm H}_2 \left( \mathbf{P}_2 \mathbf{P}^{\rm H}_2\right)^{-1} \mathbf{P}_2 + \left( \mathbf{P}^{\rm H}_2 \left( \mathbf{P}_2 \mathbf{P}^{\rm H}_2\right)^{-1} \right) \mathbf{B}\mathbf{F}\mathbf{A}\mathbf{A}^{\rm H}\mathbf{F}^{\rm H}\mathbf{B}^{\rm H}   \left( \mathbf{P}^{\rm H}_2 \left( \mathbf{P}_2 \mathbf{P}^{\rm H}_2\right)^{-1} \right)^{\rm H} \nonumber \\
&= \mathbf{I}_N - \mathbf{P}^{\rm H}_2 \left( \mathbf{P}_2 \mathbf{P}^{\rm H}_2\right)^{-1} \mathbf{P}_2 + N \left( \mathbf{P}^{\rm H}_2 \left( \mathbf{P}_2 \mathbf{P}^{\rm H}_2\right)^{-1} \right) \mathbf{B}\mathbf{B}^{\rm H}   \left( \mathbf{P}^{\rm H}_2 \left( \mathbf{P}_2 \mathbf{P}^{\rm H}_2\right)^{-1} \right)^{\rm H} \nonumber \\
&= \mathbf{I}_N - \mathbf{P}^{\rm H}_2 \left( \mathbf{P}_2 \mathbf{P}^{\rm H}_2\right)^{-1} \mathbf{P}_2 + \left( \mathbf{P}^{\rm H}_2 \left( \mathbf{P}_2 \mathbf{P}^{\rm H}_2\right)^{-1} \right) \mathbf{B}\mathbf{\Phi}\mathbf{F}\mathbf{A}\mathbf{A}^{\rm H}\mathbf{F}^{\rm H}\mathbf{\Phi}^{\rm H}\mathbf{B}^{\rm H}   \left( \mathbf{P}^{\rm H}_2 \left( \mathbf{P}_2 \mathbf{P}^{\rm H}_2\right)^{-1} \right)^{\rm H} \nonumber \\
&= \mathbf{I}_N - \mathbf{P}^{\rm H}_2 \left( \mathbf{P}_2 \mathbf{P}^{\rm H}_2\right)^{-1} \mathbf{P}_2 + \mathbf{P}^{\rm H}_2 \left( \mathbf{P}_2 \mathbf{P}^{\rm H}_2\right)^{-1}  \mathbf{P}_2 \mathbf{P}^{\rm H}_2   \left( \mathbf{P}_2 \mathbf{P}^{\rm H}_2\right)^{-1} \mathbf{P}_2    \nonumber \\
&= \mathbf{I}_N.
\end{align}

Therefore, we prove that when $\beta=0$, the smoothed data $\bar{d}_{i,k,m}$ are uncorrelated. Moreover, Eq. \eqref{Eqn62} proves that when $\beta=0$
\begin{equation}
  \label{Eqn63}
\tilde{\mathbf{P}} = \hat{\mathbf{P}} \hat{\mathbf{P}} ^{\rm H}.
\end{equation}


\begin{thebibliography}{99}

\bibitem{Ref1}
N. Michailow, M. Matth\'e, I. Gaspar, A. Navarro Caldevilla, L. L. Mendes, A. Festag, and G. Fettweis, ``Generalized frequency division multiplexing for 5th generation cellular networks," \emph{IEEE Trans. on Commun.}, vol. 62, no. 9, pp. 3045-3061, Sep. 2014.

\bibitem{Ref2}
R. Datta, N. Michailow, M. Lentmaier, and G. Fettweis, ``GFDM interference cancellation for flexible cognitive radio PHY design," in \emph{Proc. 76th IEEE VTC Fall}, Qu\'ebec City, QC, Canada, Sep. 2012, pp. 1-5.

\bibitem{Ref3}
G. Wunder, P. Jung, M. Kasparick, T. Wild, F. Schaich, Y. Chen, S. Brink, I. Gaspar, N. Michailow, A. Festag, L. Mendes, N. Cassiau, D. Ktenas, M. Dryjanski, S. Pietrzyk, B. Eged, P. Vago, and F. Wiedmann, ``5GNOW: Non-orthogonal, asynchronous waveforms for future mobile applications," \emph{IEEE Commun. Mag.}, vol. 52, no. 2, pp. 97-105, Feb. 2014.

\bibitem{Ref4}
G. Fettweis, M. Krondorf, and S. Bittner, ``GFDM--Generalized frequency division multiplexing," in {\it Proc. 69th IEEE VTC Spring}, Barcelona, Spain, Apr. 2009, pp. 1-4.

\bibitem{Ref5}
I. Gaspar, L. Mendes, M. Matth\'e, N. Michailow, D. Zhang, A. Alberti, and G. Fettweis, ``GFDM--a framework for virtual PHY services in 5G networks," arXiv:1507.04608v1, Jul. 2015.

\bibitem{Ref6}
M. Matth\'e, N. Michailow, I. Gaspar,  and G. Fettweis, ``Influence of pulse shaping on bit error rate performance and out of band radiation of generalized frequency division multiplexing," presented at the Internal Conf. Communications Workshop 5G Technologies, Sydney, NSW, Australia, Jun. 2014.

\bibitem{Ref7}
M. Matthe, L. L. Mendes, and G. Fettweis, ``Generalized frequency division multiplexing in a Gabor transform setting," \emph{IEEE Commun. Letters}, vol. 18, no. 8, pp. 1379-1382, Aug. 2014.

\bibitem{Ref8}
B. Farhang-Boroujeny, A. Farhang, A. RezazadehReyhani, A. Aminjavaheri, D. Qu, ``A comparison of linear FBMC and circularly shaped waveforms," in {\it Proc. 2016 IEEE/ACES International Conference on Wireless Information Technology and Systems (ICWITS) and Applied Computational Electromagnetics (ACES)}, Honolulu, HI, USA, Mar. 2016, pp. 1-2.





\bibitem{Ref9}
S. Brandes, I. Cosovic, and M. Schnell, ``Reduction of out-of-band radiation in OFDM systems by insertion of cancellation carriers," \emph{IEEE Commun. Lett.}, vol. 10, no. 6, pp. 420-422, Jun. 2006.

\bibitem{Ref10}
P. Kryszkiewicz and H. Bogucka, ``Out-of-band power reduction in NC-OFDM with optimized cancellation carriers selection," \emph{IEEE Commun. Lett.}, vol. 17, no. 10, pp. 1901-1904, Oct. 2013.

\bibitem{Ref11}
Z. Wang, D. Qu, T. Jiang, and Y. He, ``Spectral sculpting for OFDM based opportunistic spectrum access by extended active interference cancellation," in \emph{Proc. IEEE Global Telecommun. Conf. (GLOBECOM)}, New Orleans, USA, Dec. 2008, pp. 4442-4446.
\bibitem{Ref12}
D. Qu, Z. Wang, and T. Jiang, ``Extended active interference cancellation for sidelobe suppression in cognitive radio OFDM systems with cyclic prefix," \emph{IEEE Trans. Veh. Technol.}, vol. 59, no. 4, pp. 1689-1695, May 2010.


\bibitem{Ref13}
M. Ma, X. Huang, B. Jiao, and Y. J. Guo, ``Optimal orthogonal precoding for power leakage suppression in DFT-based systems," \emph{IEEE Trans. Commun.}, vol. 59, no. 3, pp. 844-853, Mar. 2011.
\bibitem{Ref14}
J. Zhang, X. Huang, A. Cantoni, and Y. J. Guo, ``Sidelobe suppression with orthogonal projection for multicarrier systems," \emph{IEEE Trans. Commun.}, vol. 60, no. 2, pp. 589-599, Feb. 2012.
\bibitem{Ref15}
C. D. Chung, ``Spectrally precoded OFDM," \emph{IEEE Trans. Commun.}, vol. 54, no. 12, pp. 2173-2185, Dec. 2006.


\bibitem{Ref16}
J. van de Beek and F. Berggren, ``\emph{N}-continuous OFDM," \emph{IEEE Commun. Lett.}, vol. 13, no. 1, pp. 1-3, Jan. 2009.

\bibitem{Ref17}
J. van de Beek, ``Sculpting the multicarrier spectrum: a novel projection precoder," \emph{IEEE Commun. Lett.}, vol. 13, no. 12, pp. 881-883, Dec. 2009.
\bibitem{Ref18}
J. van de Beek, ``Orthogonal multiplexing in a subspace of frequency well-localized signals," \emph{IEEE Commun. Lett.}, vol. 14, no. 10, pp. 882-884, Oct. 2010.
\bibitem{Ref19}
M. Ohta, A. Iwase, and K. Yamashita, ``Improvement of the error characteristics of an \emph{N}-continuous OFDM system with low data channels by SLM," in \emph{Proc. IEEE Int. Conf. Commun. (ICC)}, Kyoto, Japan, Jun. 2011, pp. 1-5.
\bibitem{Ref20}
M. Ohta, M. Okuno, and K. Yamashita, ``Receiver iteration reduction of an \emph{N}-continuous OFDM system with cancellation tones," in \emph{Proc. IEEE Global Telecommun. Conf. (GLOBECOM)}, Kathmandu, Nepal, Dec. 2011, pp. 1-5.



\bibitem{Ref21}
P. Wei, L. Dan, Y. Xiao, and S. Li, ``A low-complexity time-domain signal processing algorithm for {\it N}-continuous OFDM," in {\it Proc. IEEE Int. Conf. Commun. (ICC)}, Budapest, Hungary, Jun. 2013, pp. 5754-5758.

\bibitem{Ref22}
P. Wei, L. Dan, Y. Xiao, W. Xiang, and S. Li, ``Time-domain {\it N}-continuous OFDM: System architecture and performance analysis," {\it IEEE Trans. Veh. Technol.}, accepted in May 2016.

\bibitem{Ref23}
G. Strang, {\it Linear Algebra and its Applications}, 3rd ed. San Diego, CA: Harcourt Brace Jovanovich Publishers, 1976.









%\bibitem{Ref16}
%R. Datta, N. Michailow, M. Lentmaier, and G. Fettweis, ``GFDM interference cancellation for flexible cognitive radio PHY design," in \emph{Proc. 76th IEEE VTC Fall}, Qu\'ebec City, QC, Canada, Sep. 2012, pp. 1-5.
%
%\bibitem{Ref23}
%G. Wunder, P. Jung, M. Kasparick, T. Wild, F. Schaich, Y. Chen, S. Brink, I. Gaspar, N. Michailow, A. Festag, L. Mendes, N. Cassiau, D. Ktenas, M. Dryjanski, S. Pietrzyk, B. Eged, P. Vago, and F. Wiedmann, ``5GNOW: Non-orthogonal, asynchronous waveforms for future mobile applications," \emph{IEEE Commun. Mag.}, vol. 52, no. 2, pp. 97-105, Feb. 2014.
%
%\bibitem{Ref16}
%M. Matthe, L. L. Mendes, and G. Fettweis, ``Generalized frequency division multiplexing in a Gabor transform setting," \emph{IEEE Commun. Letters}, vol. 18, no. 8, pp. 1379-1382, Aug. 2014.
%
%\bibitem{Ref22}
%I. Gaspar, N. Michailow, A. Navarro, E. Ohlmer, S. Krone, and G. Fettweis, ``Low complexity GFDM receiver based on sparse frequency domain processing," in \emph{Proc. 77th IEEE VTC Spring}, Dresden, Germany, Jun. 2013, pp. 1-6.
%
%\bibitem{Ref19}
%I. Gaspar, L. Mendes, M. Matth\'e, N. Michailow, D. Zhang, A. Alberti, and G. Fettweis, ``GFDM--a framework for virtual PHY services in 5G networks," arXiv:1507.04608v1, Jul. 2015.
%
%\bibitem{Ref17}
%M. Matthe, I. Gaspar, D. Zhang, and G. Fettweis, ``Reduced complexity calculation of LMMSE filter coefficients for GFDM," arXiv:1503.02782v2, Apr. 2015.
%
%\bibitem{Ref18}
%I. Gaspar, M. Matthe, N. Michailow, L. L. Mendes, D. Zhang, and G. Fettweis, ``GFDM transceiver using precoded data and low-complexity multiplication in time domain," arXiv:1506.03350v1, Jun. 2015.
%
%\bibitem{Ref21}
%J. Wexler and S. Raz, ``Discrete Gabor expansions," \emph{IEEE Trans. Signal Process.}, vol. 21, no. 3, pp. 207-220, Nov. 1990.
%
%\bibitem{Ref22}
%S. Qian and D. Chen, ``Discrete Gabor transform," \emph{IEEE Trans. Signal Process.}, vol. 41, no. 7, pp. 2429-2438, Jul. 1993.
%
%\bibitem{Ref13}
%S. Qian, \emph{Introduction to Time-Frequency and Wavelet Transforms}. Upper Saddle River, NJ: Prentice Hall PTR, 2002.
%
%\bibitem{Ref23}
%X.-G. Xia, ``On characterization of the optimal biorthogonal window functions for Gabor transforms," \emph{IEEE Trans. Signal Process.}, vol. 44, no. 1, pp. 133-136, Jan. 1996.
%
%\bibitem{Ref6}
%X.-G. Xia, ``A family of pulse-shaping filters with ISI-free matched and unmatched filter properties," \emph{IEEE Trans. Commun.}, vol. 45, no. 10, pp. 1157-1158, Oct. 1997.
%
%\bibitem{Ref21}
%R. A. Horn and C. R. Johnson, \emph{Matrix Analysis}. Cambridge, U. K.: Cambridge Univ. Press, 1987.
%
%\bibitem{Ref23}
%W. C. Jakes, \emph{Microwave Mobile Communications}. Piscataway, NJ: IEEE Press, 1994.
%
%%\bibitem{Ref7}
%%C. Xiao, Y. R. Zheng, and N. C. Beaulieu, ``Second-order statistical properties of the WSS Jakes? fading channel simulator,"
%
%\bibitem{Ref14}
%H. Lin and P. Siohan, ``Orthogonality improved GFDM with low complexity implementation," \emph{2015 IEEE Wireless Commun. Networking Conf. (WCNC)}, New Orleans, United states, Jun. 2015, pp. 597-602.
%
%\bibitem{Ref15}
%A. Farhang, N. Marchetti, and L. E. Doyle, ``Low complexity GFDM receiver design: a new approach," \emph{ 2015 IEEE Int. Conf. Commun. (ICC)}, London, England, Jun. 2015, pp. 4775-4780.
%
%\bibitem{Ref16}
%A. Farhang, N. Marchetti, and L. E. Doyle, ``Low complexity modem design for GFDM," \emph{IEEE Trans. Signal Process.}, vol. 64, no. 6, pp. 1507-1518, Mar. 2016.
%
%\bibitem{Ref7}
%B. Farhang-Boroujeny and H. Moradi, ``Derivation of GFDM based on OFDM principles," \emph{ 2015 IEEE Int. Conf. Commun. (ICC)}, London, England, Jun. 2015, pp. 2680-2685.
%
%\bibitem{Ref160}
%Y. Zhou, J. Wang, and T. S. Ng, ``Downlink transmission of broadband OFCDM systems--Part V: Code assignment," {\it IEEE Trans. Wireless Commun.}, vol. 7, no. 11, pp. 4546-4557, Nov. 2008.
%
%\bibitem{Ref161}
%W. C. Y. Lee, {\it Mobile Communications Engineering: Theory and Applications}, 2nd edition. New York: McGraw-Hill, 1997.
%
%\bibitem{Ref162}
%B. Sklar, ``Rayleigh fading channels in mobile digital communication systems--Part I: Characterization," {\it IEEE Commun. Mag.}, vol. 35, no. 7, pp. 90-100, Jul. 1997.


\end{thebibliography}
\end{document}